\documentclass{article}

\usepackage{arxiv}

\usepackage[utf8]{inputenc} % allow utf-8 input
\usepackage[T1]{fontenc}    % use 8-bit T1 fonts
\usepackage{hyperref}       % hyperlinks
\usepackage{url}            % simple URL typesetting
\usepackage{booktabs}       % professional-quality tables
\usepackage{amsfonts}       % blackboard math symbols
\usepackage{nicefrac}       % compact symbols for 1/2, etc.
\usepackage{microtype}      % microtypography
\usepackage{lipsum}
\usepackage{amsmath}
\usepackage{amsbsy}

\makeatletter
\renewcommand*\env@matrix[1][\arraystretch]{%
  \edef\arraystretch{#1}%
  \hskip -\arraycolsep
  \let\@ifnextchar\new@ifnextchar
  \array{*\c@MaxMatrixCols c}}
\makeatother

\usepackage{graphicx}
\graphicspath{{fig/}}

\title{2-Cluster Fixed-Point Analysis of Mean-Coupled Stuart-Landau Oscillators in the Center Manifold}

\author{
  Felix P.~Kemeth \\
  Department of Chemical and Biomolecular Engineering\\
  Whiting School of Engineering, Johns Hopkins University\\
  Baltimore, MD 21218, USA\\
  \texttt{fkemeth1@jh.edu} \\
  \And
  Bernold Fiedler \\
  Institut f\"{u}r Mathematik\\
  Freie Universit\"{a}t Berlin\\
  14195 Berlin, Germany
  \And
  Sindre W. Haugland\\
  Physik-Department, Nonequilibrium Chemical Physics,\\
  Technische Universität M\"{u}nchen,\\
  85748 Garching, Germany\\
  \And
  Katharina Krischer\\
  Physik-Department, Nonequilibrium Chemical Physics,\\
  Technische Universität M\"{u}nchen,\\
  85748 Garching, Germany
}

\begin{document}
\maketitle

\begin{abstract}
  We reduce the dynamics of an ensemble of mean-coupled Stuart-Landau oscillators
  close to the synchronized solution.
  In particular, we map the system onto the center manifold of the Benjamin-Feir instability,
  the bifurcation destabilizing the synchronized oscillation.
  Using symmetry arguments, we describe the structure of the dynamics on this center manifold up to
  cubic order, and derive expressions for its parameters.
  This allows us to investigate phenomena described by the Stuart-Landau ensemble, such
  as clustering and cluster singularities, in the lower-dimensional center manifold,
  providing further insights into the symmetry-broken dynamics of coupled oscillators.
  We show that cluster singularities in the Stuart-Landau ensemble correspond to vanishing
  quadratic terms in the center manifold dynamics.
  In addition, they act as organizing centers for the saddle-node bifurcations
  creating unbalanced cluster states as well for the transverse bifurcations
  altering the cluster stability.
  Furthermore, % we discuss where 2-cluster states are stable, and
  we show that bistability of different solutions with the same cluster-size distribution can
  only occur when either cluster contains at least $1/3$ of the oscillators,
  independent of the system parameters.
\end{abstract}

% keywords can be removed

\section{Introduction}
\label{sec:int}

Long-range interactions play a crucial role in various dynamical phenomena observed in nature.
% many physical, biological and chemical systems.
In a swarm of flashing fireflies, they may act as a synchronizing force,
causing the swarm to flash in unison.
Analogously, in an audience clapping, the acoustic sound of the clapping can be recognized by each individual, leading to clapping in unison.
In these cases, long-range interactions lead to the synchronization of individual units~\cite{strogatz00_from_kuram_to_crawf}.

On the other hand, long-range interactions may also lead to a split up of the individuals
into two or more groups, also called dynamical clustering.
In electrochemistry, a stirred electrolyte or a common resistance may induce
long-range coupling,
leading to spatial clustering on the electrode~\cite{garcia-morales08_normal_form_approac_to_spatiot,kiss06_charac_clust_format_popul_global,wang00_exper_array_global_coupl_chaot_elect_oscil,instituto05_hierar_global_coupl_induc_clust,plenge05_patter_format_stiff_oscil_media,schoenleber14_patter_format_durin_oscil_photoel,kim01_contr_chemic_turbul_by_global_delay_feedb}.
In biology, this may explain
the formation of different genotypes in an otherwise
homogeneous environment~\cite{elmhirst01_thesis, stewart03_symmet_break_origin_species}.

The individual units which experience this long-range or global coupling may
be oscillatory, as in the case of flashing fireflies or in a clapping audience,
or, as in the case of sympatric speciation, stationary genotypes.
Here, we focus on the former case of oscillatory units with long-range interactions.

Clustering in oscillatory systems with long-range interactions has been
subject to theoretical investigation for many years
~\cite{okuda93_variet_gener_clust_global_coupl_oscil,nakagawa94_from_collec_oscil_to_collec,banaji02_clust_global_coupl_oscil,daido07_aging_clust_global_coupl_oscil,ku15_dynam_trans_large_system_mean}.
See also Ref.~\cite{pikovsky15_dynam_global_coupl_oscil} for a recent review on
globally coupled oscillators.
In particular when the long-range interactions are weak compared to the intrinsic
dynamics of the oscillator, it suffices to describe the phase evolution
of each unit, and the analysis greatly simplifies~\cite{kuramoto1984_chemical_osci,
  watanabe93_integ_global_coupl_oscil_array, watanabe94_const_motion_super_josep_array}.
If, however, the influence of the coupling is strong, as in the case considered here,
such a reduction is no longer feasible and the amplitude dynamics must be considered.
Our work aims to add to the theoretical understanding of clustering in this case of strong coupling.

From the view-point of symmetry, if the coupling between $N$ identical oscillators is global (i.e. all-to-all),
then the governing equations are equivariant under the symmetric group $\mathbf{S}_N$.
This means that the evolution equations $f$ commute with elements $\sigma$ from the symmetry group,
\begin{equation}
  f(\sigma x) = \sigma f(x) \hspace{.5cm} \forall \sigma \in \mathbf{S}_N.
\end{equation}
In addition, this implies that the system has a trivial solution which is
invariant under $\mathbf{S}_N$,
that is, in which all oscillators are synchronized.
Cluster states composed of two clusters, also called 2-cluster states,
can then be viewed as states with the reduced symmetry $\mathbf{S}_{N_1} \times \mathbf{S}_{N_2}$,
with $N_1$ and $N_2$ being the number of oscillators in each cluster.
Using the equivariant branching lemma,
it can then be shown that these 2-cluster states bifurcate off the
trivial solution~\cite{elmhirst01_thesis, golubitsky2002_the_symm_perspective}.
The bifurcation at which the synchronized motion becomes unstable and the 2-cluster branches
(also called primary branches) emerge is commonly referred
to as the Benjamin-Feir instability~\cite{hakim92_dynam_global_coupl_compl_ginzb_landau_equat,
  benjamin67_disin_wave_train_deep_water_part}.

The intrinsic dimensionality of each oscillatory unit may range from $d=2$
for FitzHugh-Nagumo~\cite{fitzhugh55_mathem_model_thres_phenom_nerve_membr}
and Van~der~Pol oscillators~\cite{pol26_lxxxv},
via $d=3$ for the Oregonator~\cite{field74_oscil_chemic_system}
to $d=4$ for the original
Hodgkin-Huxley model~\cite{hodgkin52_quant_descr_membr_curren_its}, and even higher
for more detailed physical models~\cite{andreozzi19_phenom_model_nav1}.
A system composed of $N$ of these oscillators thus lives in a $d \cdot N$-dimensional phase space,
making its full investigation unfeasible even for small $d$ and $N$.
One can, however, circumvent this problem of increasingly large dimensions
by restricting the dynamics to the center manifold of certain bifurcations.
In particular, it is known that the center space of the Benjamin-Feir instability
is $N-1$ dimensional~\cite{golubitsky2002_the_symm_perspective,dias06_secon_bifur_system_with_all},
and thus a reduction to the center manifold at this bifurcation
allows for reducing the dimension of the problem to $N-1$ and thus by a factor of $\approx d$.
As we show below, such a reduction lets us reveal invariant sets and bifurcation curves
analytically -- a difficult task in the original $d\cdot N$-dimensional space.
%% Globally coupled system -> S_N symmetry
%% Cluster States -> symmetry broken states.

In this work, we focus on a particular example of a globally coupled system,
in which the network is composed of oscillating units called Stuart-Landau oscillators,
each represented by a complex variable $W_k\in \mathbb{C}$.
As opposed to phase oscillators,
each Stuart-Landau oscillator has two degrees of freedom,
i. e. an amplitude and a phase.
With a linear global coupling, the dynamics are then given by
\begin{equation}
  \dot{W}_k = W_k - \left(1+i\gamma\right)\left| W_k\right|^2W_k + \left(\beta_{\textrm{r}} + i\beta_{\textrm{i}}\right)
  \left(\langle \mathbf{W} \rangle - W_k\right)
  \label{eq:sle}
\end{equation}
with the complex coupling constant $\beta_{\textrm{r}}+i\beta_{\textrm{i}}$ and the real parameter $\gamma$,
also called the shear~\cite{aronson90_amplit_respon_coupl_oscil}.
$\langle \cdot \rangle$ indicates the ensemble mean and $\dot{W}=dW/dt$.
Bold face $\mathbf{W}$ indicates a vector containing the ensemble values $\left[W_1, W_2, \dots, W_N\right]$.
For $\beta_{\textrm{r}}+i\beta_{\textrm{i}} = 0$ the ensemble is decoupled,
and each Stuart-Landau oscillator oscillates with unit amplitude and angular velocity $-\gamma$.
For $\beta_{\textrm{r}}+i\beta_{\textrm{i}} \neq 0$, however,
a plethora of different dynamical states can be observed.
These states include fully synchronized oscillations, in which all oscillators maintain an amplitude
equal to one and have a mutual phase difference of zero~\cite{pikovsky01_synchronization},
cluster states, in which the ensemble splits up into two or more sets of synchrony~\cite{roehm18_bistab_two_simpl_symmet_coupl,kemeth19_clust_singul,daido07_aging_clust_global_coupl_oscil},
and a variety of quasi-periodic and chaotic dynamics~\cite{nakagawa93_collec_chaos_popul_global_coupl_oscil,nakagawa94_from_collec_oscil_to_collec}.

\begin{figure}[ht]
  \centering
  \includegraphics{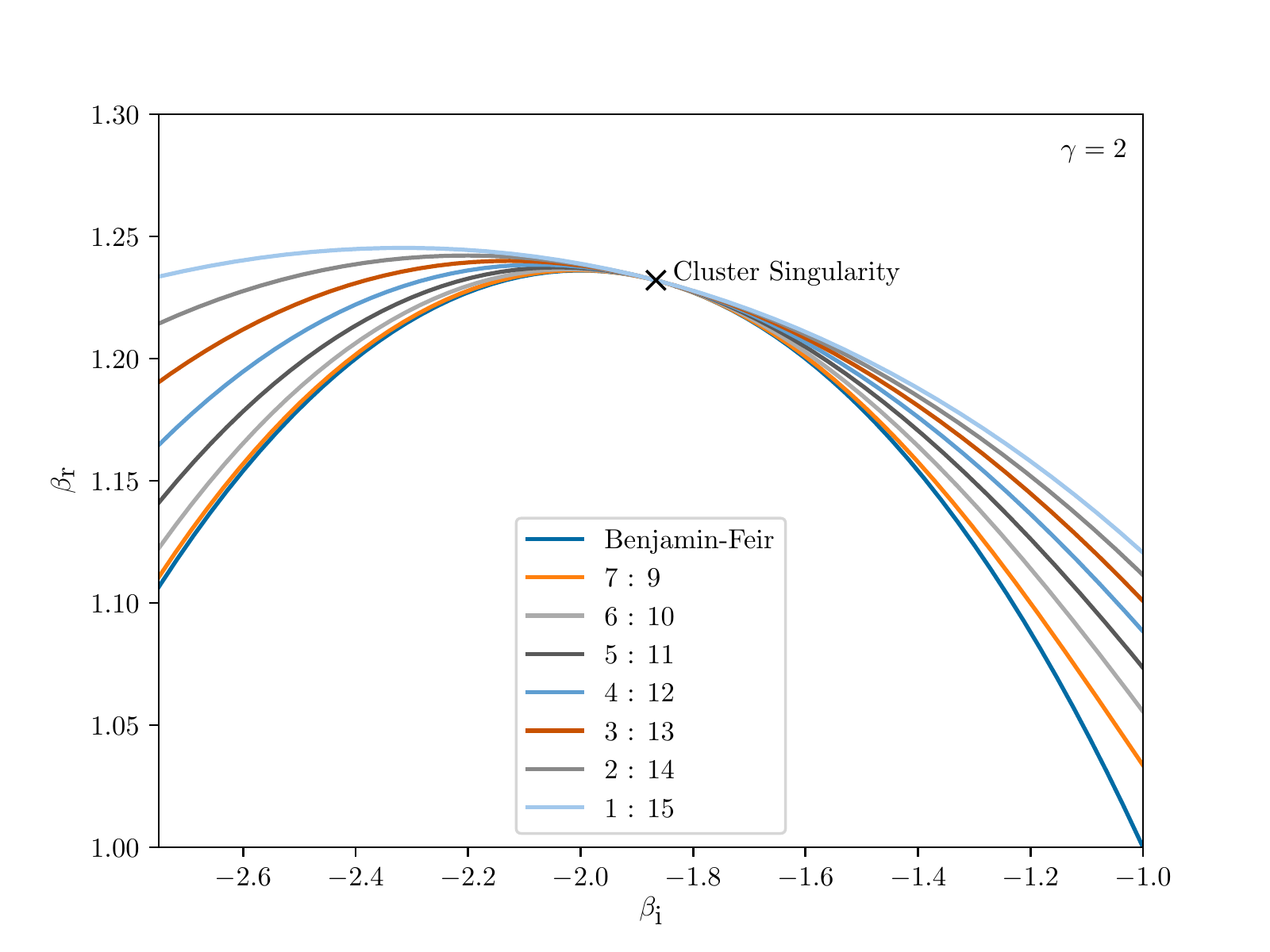}
  \caption{The Benjamin-Feir instability involving the $8:8$ cluster (dark blue) and the different saddle-node curves creating the unbalanced cluster solutions, $N_1 \neq N_2$,
    in the $\beta_{\textrm{i}}$, $\beta_{\textrm{r}}$ plane with $\gamma=2$ and $N=16$.
    Each curve belongs to a particular cluster distribution $N_1 : N_2$, and is obtained with numerical continuation using \textit{auto-07p}~\cite{doedel81_auto, doedel07_auto}.
    Note the position of the cluster singularity at $\beta_{\textrm{r}}=-(1-\sqrt{3}\gamma)/2\approx 1.23$, $\beta_{\textrm{i}}=(-\gamma-\sqrt{3})/2\approx-1.87$ as indicated.}
  \label{fig:sn_curves_sle}
\end{figure}

2-cluster states can be born and destroyed at saddle-node bifurcations if
the number of oscillators in each cluster is different, that is, when they are unbalanced
~\cite{banaji02_clust_global_coupl_oscil}.
Balanced solutions with $N_1=N_2$ emerge from the synchronized solution at
the Benjamin-Feir instability.
For $N=16$ oscillators and $\gamma=2$,
the saddle-node bifurcations for different unbalanced cluster distributions
$N_1 \neq N_2$ and the Benjamin-Feir instability are depicted in
Fig.~\ref{fig:sn_curves_sle}, as a function of the coupling parameters $\beta_{\textrm{r}}$ and $\beta_{\textrm{i}}$.
%% Here, all the two-cluster solutions exist below, that is for smaller $\beta_{\textrm{r}}$
%% values, than their respective saddle-node
%% bifurcation curve.
Here, all the 2-cluster solutions exist locally in parameter space below
their respective saddle-node
bifurcation curve, that is for
smaller $\beta_{\textrm{r}}$ values.
Up to the Benjamin-Feir instability they coexist with the stable synchronized solution.
Descending from large $\beta_{\textrm{r}}$ values,
notice that the most-unbalanced cluster state with $N_1:N_2=1:15$ is created first.
The more balanced cluster states are born subsequently, depending on their distribution,
until eventually the balanced cluster state $N_1:N_2=8:8$ is born at the Benjamin-Feir instability.
% In addition, there exists a codimension-two point %,
At $\beta_{\textrm{r}}=-(1-\sqrt{3}\gamma)/2$, $\beta_{\textrm{i}}=(-\gamma-\sqrt{3})/2$,
there exists a codimension-two point where the saddle-node bifurcations
of all cluster distributions coincide.
This point is called a \textit{cluster singularity}~\cite{kemeth19_clust_singul}.
Note that the qualitative picture in Fig.~\ref{fig:sn_curves_sle} does not
change when increasing the total number of oscillators $N$.
For large numbers $N\rightarrow \infty$
we expect a bow-tie-shaped band of saddle-node bifurcation curves,
ranging from the saddle-node bifurcation of the most unbalanced cluster state to
the Benjamin-Feir instability.
As argued in Ref.~\cite{kemeth19_clust_singul}, the cluster singularity can thus
be viewed as an organizing center.
By projecting the dynamics close to the Benjamin-Feir instability onto its center manifold,
we aim to obtain further insights into the properties of this organizing center,
and to elucidate the clustering behavior near it.

The remainder of this article is organized as follows:
In Sec.~\ref{sec:var_trans}, we pass to a corotating frame
and introduce the average amplitude $R$, the deviations from the average amplitude $r_k$,
the deviations from the mean phase $\varphi_k$.
Using this corotating system, we discuss how one can describe the dynamics in the center manifold,
see Sec.~\ref{sec:cmr}.
In Sec.~\ref{sec:ests}, we derive the parameters for the dynamics of $x_k$.
Detailed calculations are provided in Appendix~\ref{sec:par_ests}, for convenience.
Based on the parameters in the center manifold,
we study the bifurcations of 2-cluster states and the role of the cluster singularity in the
center manifold, in Sec.~\ref{sec:sle_cmf}.
We conclude with a detailed discussion of our results
and an outlook on future work.
For a detailed mathematical analysis of the dynamics of 2-cluster states in the center manifold,
see the companion paper Ref.~\cite{fiedler20_dynamics}.

\section{Variable transformation into corotating frame}
\label{sec:var_trans}
Notice that Eq.~\eqref{eq:sle} is invariant under a rotation in the complex plane $W_k \rightarrow W_k\exp(i\phi)$.
This invariance can be eliminated by choosing variables in a corotating frame,
thus effectively reducing the dimensions of the system from $2N$ to $2N-1$.

In particular, we express the complex variables $W_k$ in
log-polar coordinates $W_k=\exp(R_k+i\Phi_k)$.
Then Eq.~\eqref{eq:sle} turns into
\begin{align}
  \dot{R} & = 1-e^{2R} \langle e^{2\mathbf{r}}\rangle +
  \operatorname{Re} \left(\left(\beta_{\textrm{r}}+i\beta_{\textrm{i}}\right)\left(\langle e^{\mathbf{z}}\rangle \langle e^{-\mathbf{z}} \rangle-1\right)\right)
  \tag{3a}\label{eq:3a}\\
  \dot{r}_k & = - e^{2R}\widetilde{e^{2r_k}} +
  \operatorname{Re} \left(\left(\beta_{\textrm{r}}+i\beta_{\textrm{i}}\right)\left(\langle e^{\mathbf{z}}\rangle \widetilde{e^{-z_k}}\right)\right)
  \tag{3b}\label{eq:3b}\\
  \dot{\varphi}_k & = -\gamma e^{2R}\widetilde{e^{2r_k}} +
  \operatorname{Im} \left(\left(\beta_{\textrm{r}}+i\beta_{\textrm{i}}\right)\left(\langle e^{\mathbf{z}}\rangle \widetilde{e^{-z_k}}\right)\right)
  \tag{3c}\label{eq:3c}
\end{align}
\setcounter{equation}{3}
with $k=1, \dots, N-1$, the abbreviations shown in Tab.~\ref{tab:nots}
and the new coordinates summarized in Tab.~\ref{tab:co_trans} (see~Appendix~\ref{sec:aaa} for a derivation).
\begin{table}[h]
  \centering
  \caption{Abbreviations}
  \begin{tabular}{ll}
    \toprule
    $\langle \mathbf{x}^m \rangle = 1/N\sum_{j=1}^N x_j^m$ &
    $\widetilde{x_k^m} = x_k^m - \langle \mathbf{x}^m\rangle$ \\
    $\langle e^{\mathbf{x}} \rangle = 1/N\sum_{j=1}^N e^{x_j}$ &
    $\widetilde{e^{x_k}} = e^{x_k} - \langle e^{\mathbf{x}}\rangle$\\
    \bottomrule
  \end{tabular}
  \label{tab:nots}
\end{table}
\begin{table}[h]
  \centering
  \caption{Coordinate transformations}
  \begin{tabular}{lll}
    \toprule
    $R=\langle \mathbf{R}\rangle$ & $r_k = \widetilde{R_k}$   & $\Rightarrow \langle \mathbf{r}\rangle=0$ \\
    $\Phi = \langle \mathbf{\Phi}\rangle$ & $\varphi_k = \widetilde{\Phi_k}$ &
    $\Rightarrow \langle \boldsymbol{\varphi}\rangle=0$ \\
    & $z_k = r_k + i\varphi_k$ & $\Rightarrow \langle \mathbf{z}\rangle=0$\\
    \bottomrule
  \end{tabular}
  \label{tab:co_trans}
\end{table}
Hereby, $\widetilde{\, \cdot \,}$ symbolizes the deviation from the
ensemble mean $\langle \cdot \rangle$,
and $R$ and $\Phi$ are the ensemble mean logarithmic amplitude and phase, respectively.
The logarithmic amplitude and phase deviation of each oscillator from their averages
are $r_k$ and $\varphi_k$.
Notice that through this construction, the averages of these deviations vanish.
Furthermore, bold face of a variable, e.g. $\mathbf{x}$, symbolizes the set of the
respective ensemble variables $\left\{ x_1, x_2, \dots, x_N\right\}$.\\
To simplify notation, $r_k+i\varphi_k$ is abbreviated by the complex variable $z_k$.
The transformation into Eqs.~\eqref{eq:3a} to~\eqref{eq:3c} has the advantage that the resulting
equations are independent of the mean phase $\Phi$.
A change of $\Phi$ corresponds to a uniform phase shift of the whole ensemble in the complex plane,
which in turn means that periodic orbits in the Stuart-Landau ensemble, Eq.~\eqref{eq:sle},
correspond to stationary solutions in the transformed system, Eqs.~\eqref{eq:3a} to~\eqref{eq:3c}.
Thus, we can ignore the mean phase $\Phi$ in our subsequent analysis.\\
Synchronized oscillations correspond to $R=0$, $r_k=0$, $\Phi=-\gamma t$ and $\varphi_k=0$.
The stability of this equilibrium can be investigated using the eigenspectrum of
the Jacobian evaluated at this point.
Due to the $\mathbf{S}_N$-symmetry of the solution
and the $\mathbf{S}_N$-equivariance of the governing equations,
the Jacobian becomes block-diagonal, and thus
has a degenerate eigenvalue spectrum~\cite{hakim92_dynam_global_coupl_compl_ginzb_landau_equat,ku15_dynam_trans_large_system_mean},
see Appendix~\ref{sec:aa}:
\begin{itemize}
\item There is one singleton eigenvalue $\lambda_1=-2<0$, corresponding to an eigendirection
affecting all oscillators identically.
That is, this direction $\vec{v}_1$ shifts the amplitude of the synchronized motion but
does not alter its symmetry.
%% \begin{align*}
%% \vec{v}_1 = \begin{pmatrix}v_{1, R}\\ \vec{v}_{1, r_k} \\ \vec{v}_{1, \varphi_k}
%% \end{pmatrix} = \begin{pmatrix} 1 \\ \vec{0} \\ \vec{0}
%% \end{pmatrix}.
%% \end{align*}
\item There is the eigenvalue $\lambda_+=-1-\beta_{\textrm{r}}+\sqrt{1-\beta_{\textrm{i}}^2-2\beta_{\textrm{i}}\gamma} =: -1-\beta_{\textrm{r}}+d$ which becomes zero at the Benjamin-Feir instability and is of geometric
multiplicity $N-1$.
The corresponding directions correspond to 2-cluster states,
with each direction corresponding to one cluster distribution $N_1:N_2$.
Up to conjugacy, we arrange here the units such that the first $N_1$ oscillators correspond to the
same cluster. All 2-clusters with the same distribution but different assignments of the oscillators
then belong to the same conjugacy class.
%% \begin{align*}
%% \vec{v}_+ = \begin{pmatrix}v_{1, R}\\ \vec{v}_{1, r_k} \\ \vec{v}_{1, r_l} \\ \vec{v}_{1, \varphi_k} \\ \vec{v}_{1, \varphi_l}
%% \end{pmatrix} = \begin{pmatrix} 0 \\ \vec{0} \\ \vec{0}
%% \end{pmatrix}.
%% \end{align*}

\item Finally, there is the eigenvalue $\lambda_-= -1-\beta_{\textrm{r}}-d$ which is negative close to the
synchronized solution, which has a geometric multiplicity of $N-1$ and whose eigendirections also
have $\mathbf{S}_{N_1}\times\mathbf{S}_{N_2}$-symmetry.\\
\end{itemize}
Hereby, $d=\sqrt{1-\beta_{\textrm{i}}^2-2\beta_{\textrm{i}}\gamma}$ abbreviates the root of the discriminant where we
assume $1-\beta_{\textrm{i}}^2-2\beta_{\textrm{i}}\gamma>0$, i.e. real $\lambda_\pm$.
Notice that the Benjamin-Feir instability $\lambda_+=0$, alias $\beta_{\textrm{r}}=d-1$, i.e. the dark blue curve in
Fig.~\ref{fig:sn_curves_sle}, is of codimension one.

\section{Center manifold reduction}
\label{sec:cmr}
In the following, we calculate an expansion to third order of the dynamics in the $(N-1)$-dimensional
center manifold which corresponds to the Benjamin-Feir instability at $\lambda_+=0=-1-\beta_{\textrm{r}}+d$.
In order to do so, it is useful to introduce the coordinates
\begin{align}
  x_k & = \frac{-r_k + \frac{d+1}{\gamma'}\varphi_k}{2d}\\
  y_k & = \frac{r_k + \frac{d-1}{\gamma'}\varphi_k}{2d}
\end{align}
such that
\begin{align}
  r_k & = \left(1-d\right)x_k + \left(1+d\right)y_k\\
  \varphi_k & = \gamma'x_k + \gamma'y_k.
\end{align}
Here we use the notations $\gamma' = 2\gamma+\beta_{\textrm{i}}$ and $d$ as defined above.
See Appendix~\ref{sec:aa} for a derivation.
The variables $x_k$ describe the dynamics in the $(N-1)$-dimensional center manifold tangent to $y_k=0 \, \forall k$, while
$y_k$ together with $R$ describe the dynamics in the stable manifold tangent to $x_k=0\,\forall k$.\\
Note that the center-manifold must be $\mathbf{S}_N$-invariant.
In addition, the global restrictions $\langle \mathbf{r}\rangle =\langle\boldsymbol{\varphi}\rangle=0$
and thus $\langle \mathbf{x}\rangle =\langle \mathbf{y}\rangle=0$ must hold.
Therefore, the general form of the center manifold up to quadratic order must follow
\begin{align}
  y_k & = y_k\left(\mathbf{x}\right) = a\widetilde{x_k^2} +\mathcal{O}\left(x_k^3\right)\label{eq:yk}\\
  R & = R\left(\mathbf{x}\right) = b \langle \mathbf{x}^2\rangle + \mathcal{O}\left(x_k^3\right)\label{eq:R}
\end{align}
with the coefficients $a=a\left(\beta_{\textrm{i}}, \gamma\right)$ and $b=b\left(\beta_{\textrm{i}}, \gamma\right)$.
Here, we use the tangency of our coordinates $R$ and $y_k$, that is,
$\left. \frac{\mathrm{d}}{\mathrm{d}x_k} R\right|_{\mathbf{x}=0}=0$ and $\left. \frac{\mathrm{d}}{\mathrm{d}x_k} y_k\right|_{\mathbf{x}=0}=0$.
Since the Benjamin-Feir instability $\beta_{\textrm{r}}=d-1$ is of codimension one,
the three-dimensional parameter space $(\beta_{\textrm{r}}, \beta_{\textrm{i}}, \gamma)$ becomes two-dimensional.
The parameters in the center manifold thus only depend on $\beta_{\textrm{i}}$ and $\gamma$.
By $\mathbf{S}_N$-equivariance, the reduced dynamics $\dot{x}_k$ in the center manifold,
up to cubic order, must be of the form
\begin{equation}
  \dot{x}_k = \lambda_+ x_k + A\widetilde{x_k^2} + B\widetilde{x_k^3} + C\langle \mathbf{x}^2\rangle x_k
  + \mathcal{O}\left(x_k^4\right),
  \label{eq:cmf}
\end{equation}
see also Refs.~\cite{stewart03_symmet_break_origin_species, golubitsky2002_the_symm_perspective}, with the parameters $A=A\left(\beta_{\textrm{i}}, \gamma\right)$ and $B=B\left(\beta_{\textrm{i}}, \gamma\right)$ and $C=C\left(\beta_{\textrm{i}}, \gamma\right)$.

\section{Derivation of the parameters $a$, $b$, $A$, $B$ and $C$}
\label{sec:ests}

In this section, we discuss the approach to calculate the coefficients $a$, $b$, $A$, $B$ and $C$
for the dynamics in the center manifold.
See Appendix~\ref{sec:par_ests} for complete details.\\
First, we determine $b$.
In particular we observe that
\begin{equation*}
  \dot{R} = \left(\frac{\mathrm{d}}{\mathrm{d}x_k} R\right) \, \dot{\mathbf{x}} =
  2b \langle \mathbf{x}\dot{\mathbf{x}}\rangle +\mathcal{O}\left(x_k^5\right) =
  2b\lambda_+ \langle \mathbf{x}^2 \rangle+
  \mathcal{O}\left(x_k^3\right)
\end{equation*}
holds. Since $\lambda_+=0$ at the bifurcation, $\dot{R}$ up to second order in $x_k$
must vanish.
Therefore, expressing $z_k=r_k + i \varphi_k$ and $r_k$, $\varphi_k$ in terms of $x_k$
in Eq.~\eqref{eq:3a},
we can compute $b$ by comparing the coefficients of the $\langle \mathbf{x}^2\rangle$:
the terms in front of $\langle \mathbf{x}^2\rangle$ must thereby vanish.
%% The terms in front of $\langle \mathbf{x}^2\rangle$ and $b\langle \mathbf{x}^2\rangle$ must
%% thereby vanish.
This allows us to estimate $b=b\left(\beta_{\textrm{i}}, \gamma\right)$ as
\begin{equation}
  b = \frac{1-d}{2}\left(\gamma'^2 +d^2+4d-5\right)
  \label{eq:b}
\end{equation}
with $\gamma'$ and $d$ as defined above.\\
Analogously, we can calculate $a$ using Eqs.~\ref{eq:3a} and~\ref{eq:3b} up to second order in $x_k$ and employing
\begin{equation*}
  \dot{y}_k = \left(\frac{\mathrm{d}}{\mathrm{d}x_k} y_k\right) \dot{x}_k = \mathcal{O}\left(x_k^3\right).
\end{equation*}
This means we can use $2d\dot{y}_k=\dot{r}_k + (d-1)/\gamma' \dot{\varphi}_k$,
substitute the $z_k$ with $x_k$ in Eqs.~\ref{eq:3a} and~\ref{eq:3b} and keep terms up to
$\mathcal{O}\left(x_k^2\right)$.
Comparing the coefficients in front of $\widetilde{x_k^2}$ then results in
\begin{equation}
  a = \frac{\left(1-d\right)\left(\gamma'^2+\left(1-d\right)^2\right)\left(3\left(d^2-1\right)
      +\gamma'^2\right)}{8d^2\gamma'^2}.
\end{equation}
Finally, we can calculate $A$, $B$ and $C$ using
\begin{align*}
  2d\dot{x}_k & = -\dot{r}_k + \frac{d+1}{\gamma'} \dot{\varphi}_k\\
  & = \lambda_+ x_k + A\widetilde{x_k^2} + B\widetilde{x_k^3} + C\langle \mathbf{x}^2\rangle x_k.
\end{align*}
Taking Eqs.~\ref{eq:3b} and~\ref{eq:3c} and the coefficients $a$ and $b$ obtained above,
we can evaluate this equality up to cubic order, yielding the coefficients
\begin{align}
  A &= \frac{\left(d-1\right)\left(\gamma'^2+\left(1+d\right)^2\right)\left(\gamma'^2-3\left(d-1\right)^2\right)}{4\gamma'^2d}\label{eq:A}\\
  B &= -\frac{\left(d-1\right)^2\left(\gamma'^2+\left(d-1\right)^2\right)\left(\gamma'^2+\left(d+1\right)^2\right)\left(\gamma'^2-2\gamma'd+3\left(d^2-1\right)\right)\left(\gamma'^2+2\gamma'd+3\left(d^2-1\right)\right)}{16\gamma'^4d^3}\\
  C &= \frac{\left(d-1\right)^2}{16d^3\gamma'^4}\left(\vphantom{\int_1^2} 
  \gamma'^8 -4\gamma'^6\left(2d^3-7d^2+1\right)-2\gamma'^4\left(8d^5+d^4-56d^3+22d^2+1\right)\right. \nonumber\\
  & \left.-4\gamma'^2\left(2d^7+5d^6-4d^5-13d^4+2d^3+11d^2-3\right)+9\left(d^2-1\right)^4
  \vphantom{\int_1^2} \right).
\end{align}
Together with $\lambda_+$, the expressions for $A$, $B$ and $C$ fully specify the dynamics in the
center manifold based on the original parameters $\gamma$, $\beta_{\textrm{r}}$ and $\beta_{\textrm{i}}$.
By rescaling time and $x_k$ in Eq.~\eqref{eq:cmf}, the number of independent parameters
can be reduced to two, see Ref.~\cite{fiedler20_dynamics}.
For simplicity, we use the unscaled equation as in Eq.~\eqref{eq:cmf} here.

\section{Clustering and cluster singularities in the center manifold}
\label{sec:sle_cmf}

\begin{figure}[ht]
  \centering
  \includegraphics{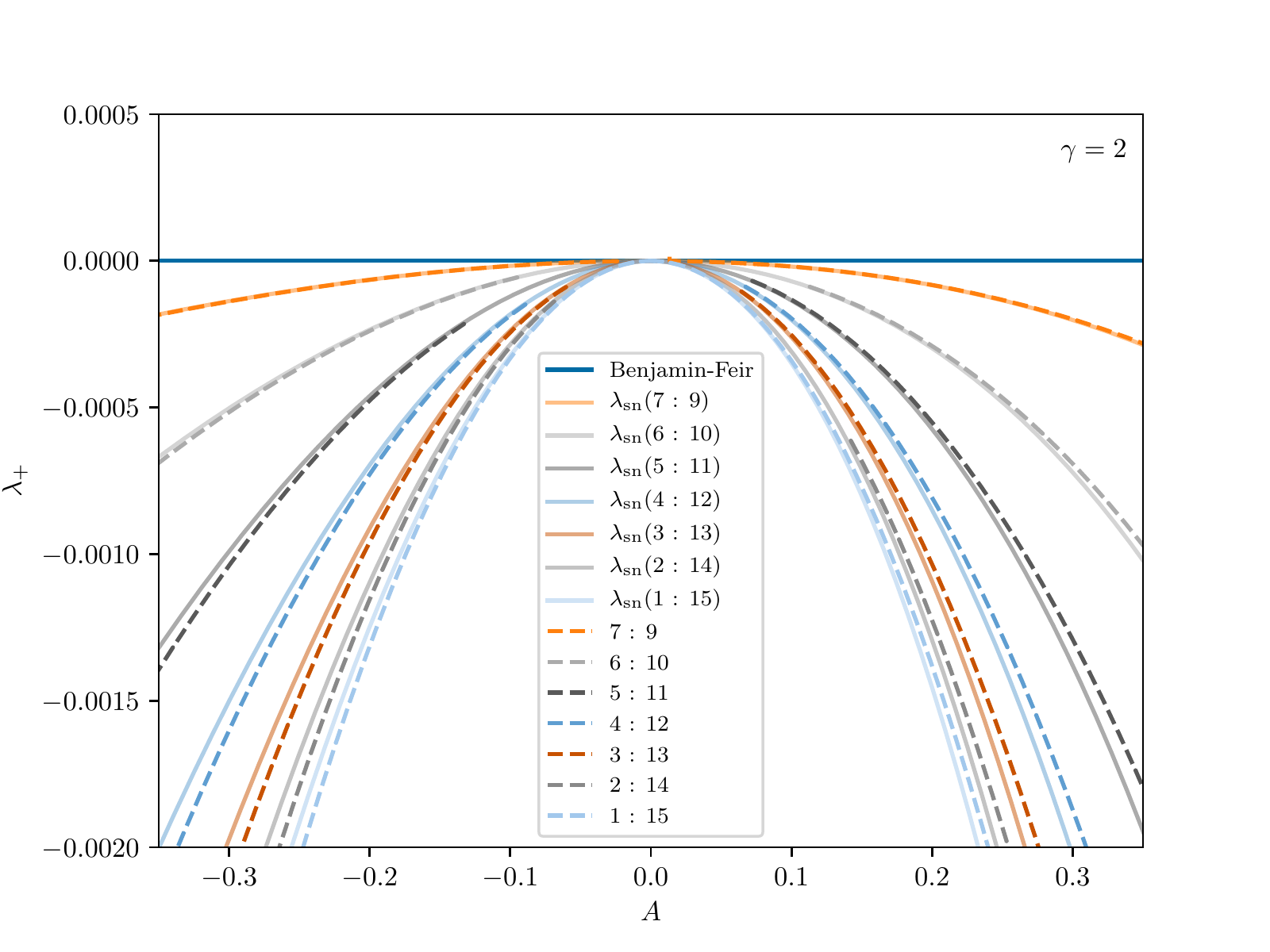}
  \caption{The Benjamin-Feir instability (blue, $\lambda_+=0$) and the
    different saddle-node curves creating the unbalanced cluster solutions in the
    $A$, $\lambda_+$ plane.
    The dashed curves belong to particular cluster distributions $N_1:N_2$
    obtained by projecting the curves from the Stuart-Landau ensemble shown in
    Fig.~\ref{fig:sn_curves_sle} using the expressions for $\lambda_+(\beta_{\textrm{r}}, \beta_{\textrm{i}}, \gamma)= -1-\beta_{\textrm{r}}+d$, and
    $A(\beta_{\textrm{r}}, \beta_{\textrm{i}}, \gamma)$, cf. Eq.~\eqref{eq:A}.
    The solid curves $\lambda_{\text{sn}}$ indicate the saddle-node
    bifurcations of the unbalanced cluster states obtained analytically
    in the center manifold, see Eq.~\eqref{eq:lambda_sn}.
    Note that close to the cluster singularity,
    where analytical expansions work best,
    numerical continuation fails
    due to the concentration of solutions in phase space.}
  \label{fig:sn_curves_cmf}
\end{figure}

As shown in Fig.~\ref{fig:sn_curves_sle} for $N=16$ oscillators,
we observe a range of saddle-node bifurcations creating the different
2-cluster states.
The expressions for $\lambda_+$, $A$, $B$ and $C$ above determine the
corresponding parameter values in the center manifold.
The respective $\lambda_+$ and $A$ values for the numerical curves shown
in Fig.~\ref{fig:sn_curves_sle}
are depicted in Fig.~\ref{fig:sn_curves_cmf} as dashed curves.
Notice that the Benjamin-Feir curve corresponds to the line $\lambda_+=0$.
Furthermore, we can derive the saddle-node curves creating unbalanced 2-cluster
states in the center manifold analytically, see Appendix~\ref{sec:twocl_cmf}.
In particular,
\begin{equation}
  \lambda_{\text{sn}} = \frac{A^2\left(1-\alpha\right)^2}{4\left(B\left(1-\alpha+\alpha^2\right)+ C\alpha\right)}
  \label{eq:lambda_sn}
\end{equation}
for unbalanced cluster solutions, with $\alpha=N_1/N_2$.
The respective analytical curves for $N=16$ are shown as solid curves in Fig.~\ref{fig:sn_curves_cmf}.
Notice the close correspondence between the mapped bifurcation curves from the full system
and the bifurcation curves determined in the center manifold.
%, in particular for more unbalanced solutions.
For less balanced solutions,
the saddle-node curves obtained from the Stuart-Landau ensemble depart more strongly
from the saddle-node curves calculated analytically in the center manifold.
We expect this to be due to
% the increased distance of these curves from the Benjamin-Feir instability and
the cubic truncation of the flow in the center manifold, thus limiting
its accuracy away from the Benjamin-Feir curve.

%% Note that for all diagrams shown in this paper,
%% we fix $\gamma=2$ and vary $\beta_{\textrm{i}}$, $\beta_{\textrm{r}}$.
%% We then used the expressions for $A(\beta_{\textrm{i}}, \beta_{\textrm{r}})$, $B(\beta_{\textrm{i}}, \beta_{\textrm{r}})$ and
%% $C(\beta_{\textrm{i}}, \beta_{\textrm{r}})$ to get the parameters in the center manifold.
%% Thus the parameters $A$, $B$ and $C$ lie on a two-dimensional manifold.
%% For the curves shown in Fig.~\ref{fig:sn_curves_cmf}, we furthermore use Eq.~\eqref{eq:lambda_sn},
%% yielding one-dimensional curves.
%% The curves are, however, not exactly parabolas, since $B$ and $C$ vary in addition to $A$,
%% which is not shown in Fig.~\ref{fig:sn_curves_cmf}.
Note that to obtain the curves in Fig.~\ref{fig:sn_curves_cmf},
we fix $\gamma=2$ and vary $\beta_{\textrm{i}}$, $\beta_{\textrm{r}}$.
We then use the expressions for $A(\beta_{\textrm{i}}, \beta_{\textrm{r}})$, $B(\beta_{\textrm{i}}, \beta_{\textrm{r}})$ and
$C(\beta_{\textrm{i}}, \beta_{\textrm{r}})$ to get the parameters in the center manifold.
Thus the parameters $A$, $B$ and $C$ lie on a two-dimensional manifold.
For the curves shown in Fig.~\ref{fig:sn_curves_cmf}, we furthermore use Eq.~\eqref{eq:lambda_sn},
yielding one-dimensional curves.
The curves are, however, not exactly parabolas, since $B$ and $C$ vary in addition to $A$,
which is not shown in Fig.~\ref{fig:sn_curves_cmf}.
For all subsequent figures, we use the values of $C=-1$ and $B=-2/(2\sqrt{3}-3)$ at the cluster
singularity for $\gamma=2$, which can be obtained analytically.
See Ref.~\cite{fiedler20_dynamics} p. 36 for a derivation.

Furthermore, from Fig.~\ref{fig:sn_curves_cmf} we observe that $A=0$, in addition to $\lambda_+=0$,
at the cluster singularity.
This means that this codimension-two point is distinguished by vanishing
quadratic dynamics in the center manifold, cf. Eq.~\ref{eq:cmf}.
In addition, it serves as an organizing center for the saddle-node bifurcations of the unbalanced cluster states:
At the saddle-node bifurcation, we have in the center manifold for a cluster state
\begin{equation*}
x_1^{\ast} = -\frac{A\left(1-\alpha\right)}{2\left(B\left(1-\alpha+\alpha^2\right)+ C\alpha\right)},
\end{equation*}
with $B<0$ and $C<0$ for the range of $\beta_{\textrm{i}}$, $\beta_{\textrm{r}}$ considered here
(not shown), see Appendix~\ref{sec:twocl_cmf}.
This means that for negative $A$ values, the saddle-node curves occur at positive $x_1$,
for positive $A$ values at negative $x_1$, and for $A=0$, at the cluster singularity, all saddle-node
bifurcations
occur at the synchronized solution $x_k=0$.
This behavior can indeed be observed in the Stuart-Landau ensemble, see Fig.~6 of
Ref.~\cite{kemeth19_clust_singul}.
%% In addition, $x_1^{\ast}$ is monotonic in $\alpha$,
%% that is, $\partial x_1^{\ast}/\partial \alpha < 0 \, \forall A \, \forall B, C<0$.
%% This means the values of saddle nodes $x_1^{\ast}$ increase with decreasing $\alpha$, that is,
%% with increasing discrepancy of the number of oscillators in each cluster.

%% \section{Stability}
%% \label{sec:stability}

The unbalanced cluster states do, in general, not emerge as stable states
from the saddle-node bifurcations.
Rather, one of the two branches created at the saddle-node bifurcation
is subsequently stabilized through transverse bifurcations
involving 3-cluster solutions with symmetry
$\mathbf{S}_{N_1}\times\mathbf{S}_{N_2}\times\mathbf{S}_{N_3}$,
also called \textit{secondary branches}~\cite{stewart03_symmet_break_origin_species}.
For a more detailed discussion on secondary branches, see also
Refs.~\cite{dias03_secon_bifur_system_with_all, dias06_secon_bifur_system_with_all}\\
In order to explain this in more detail, we follow Ref.~\cite{dias03_secon_bifur_system_with_all} Section 4.
Note that each $N_1$ : $N_2$ 2-cluster solution is invariant under
the action of the group $\mathbf{S}_{N_1}\times\mathbf{S}_{N_2}$.
From this, it follows that one can block-diagonalise the Jacobian at the 2-cluster solutions $\mathbf{S}_{N_1}\times\mathbf{S}_{N_2}$.
In doing so, one can calculate the $(N_1-1)$-degenerate eigenvalue $\mu_1$
describing the intrinsic stability of cluster $\Xi_1$, that is its stability
against transverse perturbations.
Note, however, that a cluster of size 1 cannot be broken up.
%% $N_1$ needs to be larger than 1 to capture
%% transverse stability properties of cluster $\Xi_1$.
Following Ref.~\cite{stewart03_symmet_break_origin_species} p. 23 and using isotypic decomposition,
the eigenvalue $\mu_1$ can be expressed as
\begin{equation*}
\mu_1 = \left.\mathbf{J}_{11}\right|_{\Xi_1}-\left.\mathbf{J}_{12}\right|_{\Xi_1}.
\end{equation*}
Here, $\left.\mathbf{J}_{ij}\right|_{\Xi_1}$ denotes $\partial f_i/\partial x_j$, with the respective $x_i$ and $x_j$ in cluster $\Xi_1$ and $f_i$ being the right hand side of Eq.~\eqref{eq:cmf}.
Without loss of generality, we assume in the following that $\Xi_1$ is the cluster with the smaller
number of oscillators, that is, $N_1\leq N_2$ or $\alpha\leq 1$.
Evaluating the Jacobian, one obtains that the eigenvalue $\mu_1$ changes sign at
\begin{equation}
  \lambda_{+,1}= \frac{\left(1-2\alpha\right)B- \alpha C}{\left(\alpha-2\right)^2B^2}A^2.
  \label{eq:lambda_1}
\end{equation}
Analogously, the transverse stability of cluster $\Xi_2$ is described by
\begin{equation*}
\mu_2=\left.\mathbf{J}_{11}\right|_{\Xi_2}-\left.\mathbf{J}_{12}\right|_{\Xi_2},
\end{equation*}
which changes sign at
\begin{equation}
  \lambda_{+,2} = \frac{\left(\alpha-2\right)B- C}{\left(4\alpha^2-4\alpha+1\right)B^2}\alpha A^2.
  \label{eq:lambda_2}
\end{equation}
Hereby, $\mu_2$ describes the intrinsic stability of cluster $\Xi_2$.
Furthermore notice that for the balanced cluster, $\alpha=1$ and therefore $\lambda_{+,1}=\lambda_{+,2}$.
Since both clusters contain an equal number of units, their respective intrinsic stabilities change
simultaneously.
\begin{figure}[ht]
  \centering
  \includegraphics{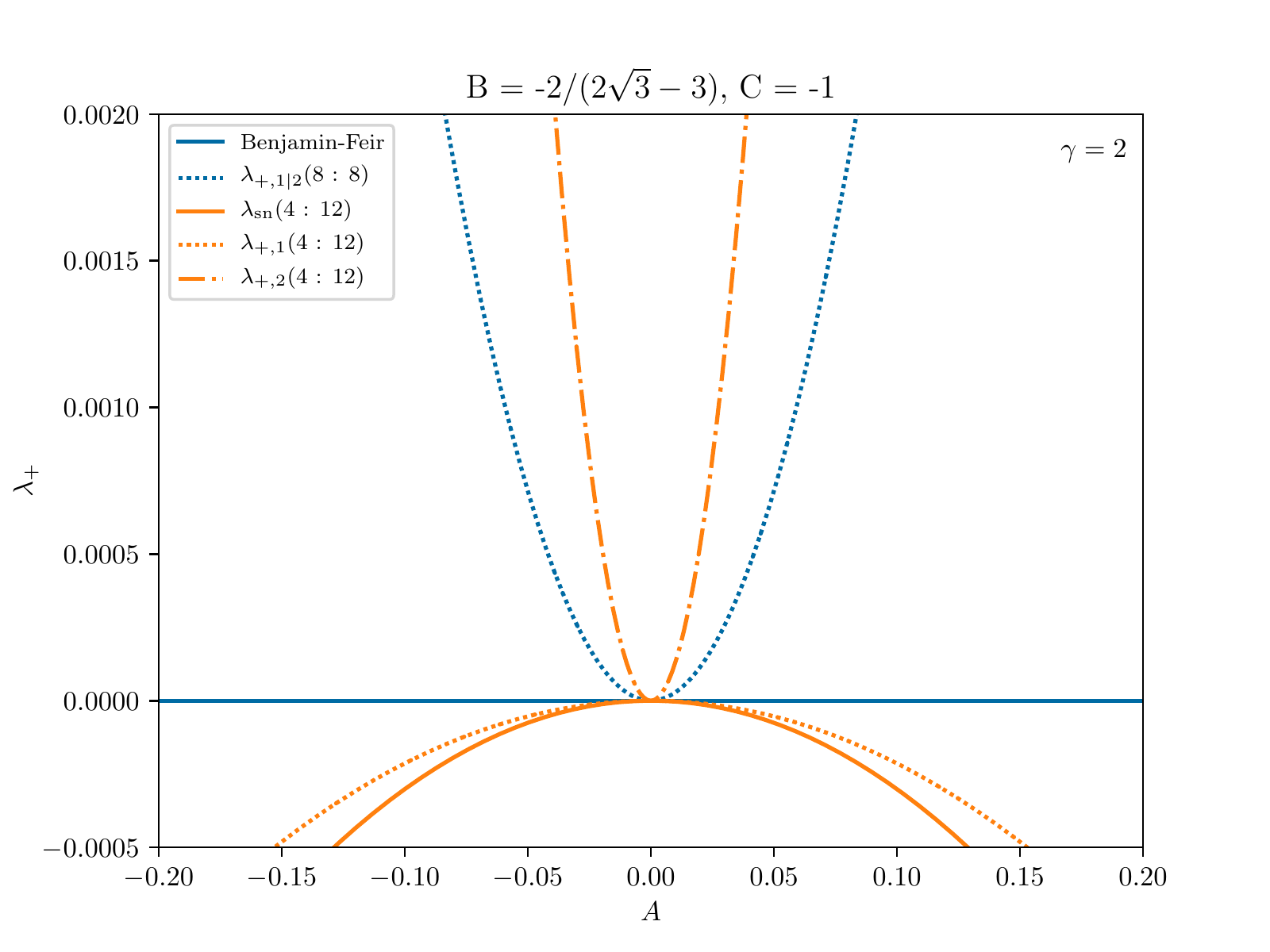}
  \caption{The bifurcation curves $\lambda_{+,1}$ ($\mu_1=0$, dotted orange) and
    $\lambda_{+,2}$ ($\mu_2=0$, dash-dotted orange)
    for the $4:12$ cluster state in the $A$, $\lambda_+$ plane and the
    parameters $B=-2/(2\sqrt{3}-3)$, $C=-1$.
    The saddle-node curve creating the $4:12$ cluster is shown as a solid orange curve.
    The Benjamin-Feir line is shown in blue, with the $\lambda_{+,1}=\lambda_{+,2}$ curve for the
    balanced $8:8$ cluster state depicted as a dotted blue curve.
    The $4:12$ cluster is stable in the two regions between the respective
    $\lambda_{+,1}=0$ and $\lambda_{+,2}=0$ curve.
    The balanced cluster state is stable above the dotted blue curve.}
  \label{fig:l1l2_79}
\end{figure}

\begin{figure}[ht]
  \centering
  \includegraphics{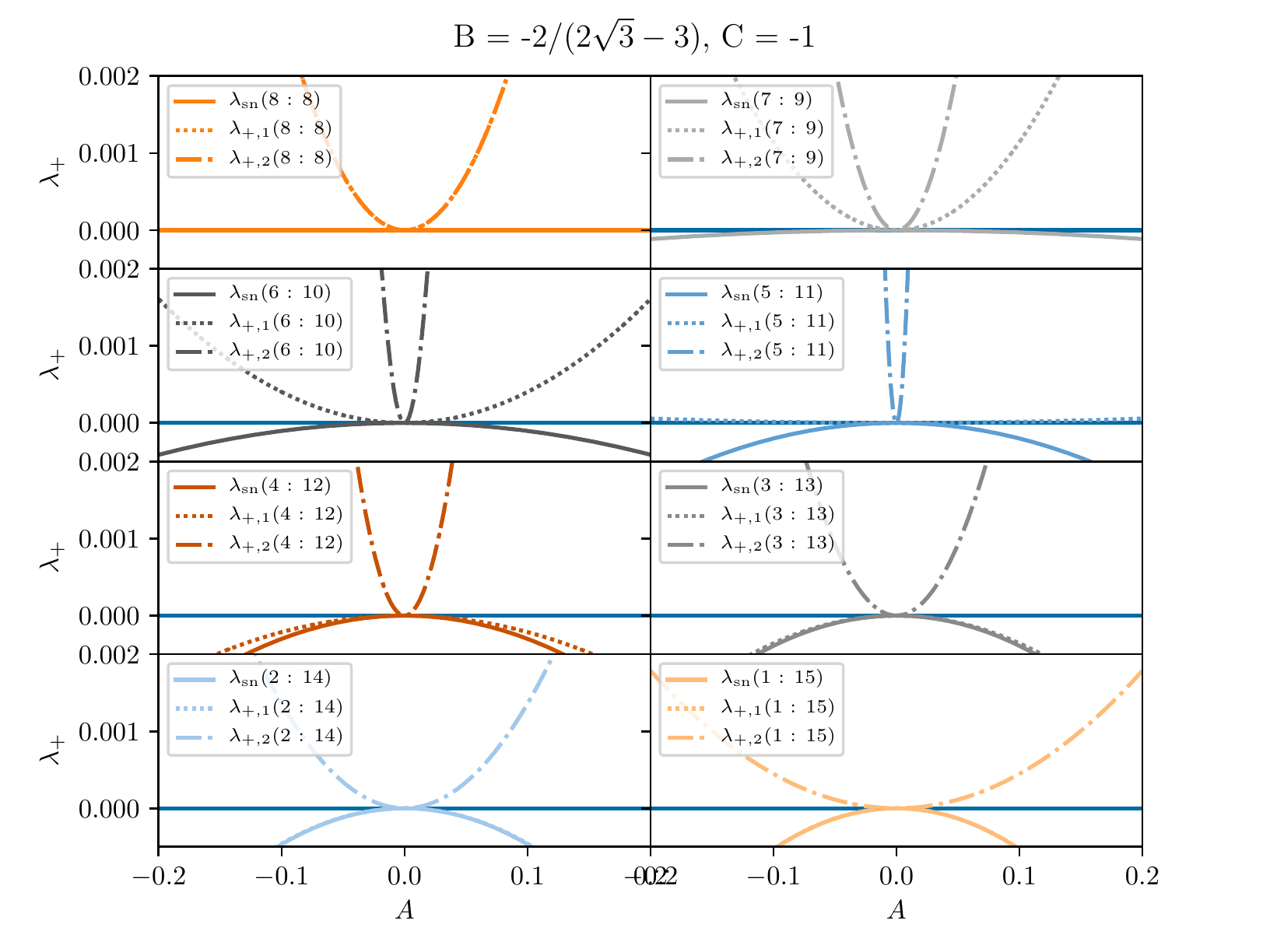}
  \caption{The theoretical bifurcation curves $\lambda_{+,1}$ ($\mu_1=0$, dotted)
    and $\lambda_{+,2}$ ($\mu_2=0$, dash-dotted)
    for the different cluster size distributions in the $A$, $\lambda_+$ plane and
    the parameters $B=-2/(2\sqrt{3}-3)$, $C=-1$.
    The saddle-node curves creating the unbalanced cluster solutions are represented
    as solid curves, which correspond to the shaded curves in Fig.~\ref{fig:sn_curves_cmf}
    with the same color coding.
    The Benjamin-Feir line is shown in blue.
    The unbalanced cluster states are stable above the
    respective dotted curve and below the dash-dotted curve, except for the $1:15$ cluster,
    which is stable already at the saddle-node bifurcation.
    For the $2:14$ cluster, the dotted and solid curves do not coincide but
    lie very close in parameter space.}
  \label{fig:l1l2}
\end{figure}

In Fig.~\ref{fig:l1l2_79}, $\lambda_{\text{sn}}$, $\lambda_{+,1}$ and $\lambda_{+,2}$ are shown as solid, dotted and dash-dotted orange curves, respectively, for the $4:12$ 2-cluster state.
The Benjamin-Feir instability, where the balanced cluster state is born, is drawn as a solid blue line at $\lambda_+=0$, and the transverse bifurcation curve $\lambda_{+,1}=\lambda_{+,2}$,
where the balanced cluster state is stabilized, is drawn as a dotted blue curve.
%% In Fig.~\ref{fig:l1l2_79}, the $\lambda_{+,1}$ (dotted orange) and $\lambda_{+,2}$ (dash-dotted orange) curves for the $4:12$ 2-cluster state are shown.
%% In addition the Benjamin-Feir instability where the balanced cluster state is born (solid blue line at $\lambda_+=0$) and the transverse bifurcation curve $\lambda_{+,1}=\lambda_{+,2}$ (dotted blue)
%% where the balanced cluster state is stabilized are drawn.
See Fig.~\ref{fig:l1l2} for the respective curves for a range of cluster distributions.

Fig.~\ref{fig:l1l2_79} can be interpreted as follows:
Coming from negative $\lambda_+$ values, the unbalanced $4:12$ cluster state is born
at $\lambda_{\text{sn}}(4:12)$ (solid orange).
However, this 2-cluster state is unstable for the parameter values considered here:
the cluster $\Xi_1$ with $4$ units is intrinsically unstable with $\mu_1>0$ and $\mu_2<0$.
At the dotted orange curve, $\mu_1$ changes sign, rendering the $4:12$ cluster state stable.
Subsequently, at the dash-dotted orange curve, $\mu_2$ changes sign,
leaving the cluster $\Xi_2$ with 12 units intrinsically unstable and thus
the $4:12$ cluster unstable.\\
The qualitatively same behavior can be observed for any cluster distribution $\alpha<1/2$, cf. Fig.~\ref{fig:l1l2},
except for the most unbalanced state ($1:15$).
There, cluster $\Xi_1$ cannot be intrinsically unstable, since it contains only one unit.
This means that this cluster solution is born stable in its saddle-node bifurcation, and becomes
unstable only at $\lambda_2$ when $\mu_2=0$. See also the bottom right plot in Fig.~\ref{fig:l1l2}.
In particular, $\lambda_{\text{sn}}=A^2/4B$ for $\alpha=0$, see Eq.~\eqref{eq:lambda_sn},
coincides with $\lambda_{+,1}=A^2/4B$ for $\alpha=0$, cf. Eq.~\eqref{eq:lambda_1}.
Furthermore, it is worth noting that the stable patches in parameter space overlap for different cluster distributions. This means that there is a multistability of different 2-cluster states.\\
Notice that these results are in close correspondence with
the behavior observed in the full Stuart-Landau ensemble, compare, for example, Fig.~\ref{fig:l1l2_79}
with Figs.~4b and~5b in Ref.~\cite{kemeth19_clust_singul}.\\
$\lambda_{\text{sn}}$ and $\lambda_{+,1}$ are continuous functions of $\alpha$.
For $N\rightarrow \infty$, this means that there are continuous bands of bifurcation curves:
Going from $\lambda_{\text{sn}}(\alpha=0)=A^2/4B$ to $\lambda_{\text{sn}}(\alpha=1)=0$,
there is band of saddle-node bifurcations creating the unbalanced cluster solutions.
This band becomes infinitesimally thin at the cluster singularity $A=0$,
giving it a bow-tie like shape.
From $\lambda_{+,1}(\alpha=0)=A^2/4B$ to $\lambda_{+,1}(\alpha=1)=(-B-C)A^2/B^2$,
the transverse bifurcations of the smaller cluster stretch from the saddle-node curve of the most unbalanced cluster state to the transverse bifurcations of the balanced cluster state where $\lambda_{+,1}$ is maximal, again yielding
a bow-tie like shape in the $A$, $\lambda_+$ plane.
Since $\lambda_{+,2}$ has a pole at $\alpha=1/2$, the interpretation is a bit more involved.
First, for the balanced cluster state $\alpha=1$:
\begin{equation*}
  \lambda_{+,2}(\alpha=1)=(-B-C)A^2/B^2=\lambda_{+,1}(\alpha=1),
\end{equation*}
and thus $\lambda_{+,2}$ and $\lambda_{+,1}$ coincide.
For the most unbalanced solution $\alpha=0$: $\lambda_{+,2}(\alpha=0)=0$.
This means the larger cluster of the most unbalanced solution becomes unstable
exactly when the balanced solution is born, that is, at the Benjamin-Feir instability $\lambda_+=0$.
For intermediate $\alpha$ values, however, the $\lambda_{+,2}$ curve becomes steeper and infinitely steep
at $\alpha=1/2$, with the tip reaching to the cluster singularity.
This can also be observed in Fig.~\ref{fig:l1l2}, where the parabola becomes
thinner when going from the $6:10$ to the $5:11$ cluster states, and subsequently broadens again until the $1:15$ cluster.
Altogether, the $\lambda_{+,2}$ curves fill out the half plane $\lambda_+\geq 0$ except the line $A=0$.\\
These three bow-tie like regions of $\lambda_{\text{sn}}$, $\lambda_{+,1}$ and $\lambda_{+,2}$ become infinitesimally thin and thus singular only at the cluster singularity $\lambda_+=0$, $A=0$.

\begin{figure}[ht]
  \centering
  \includegraphics{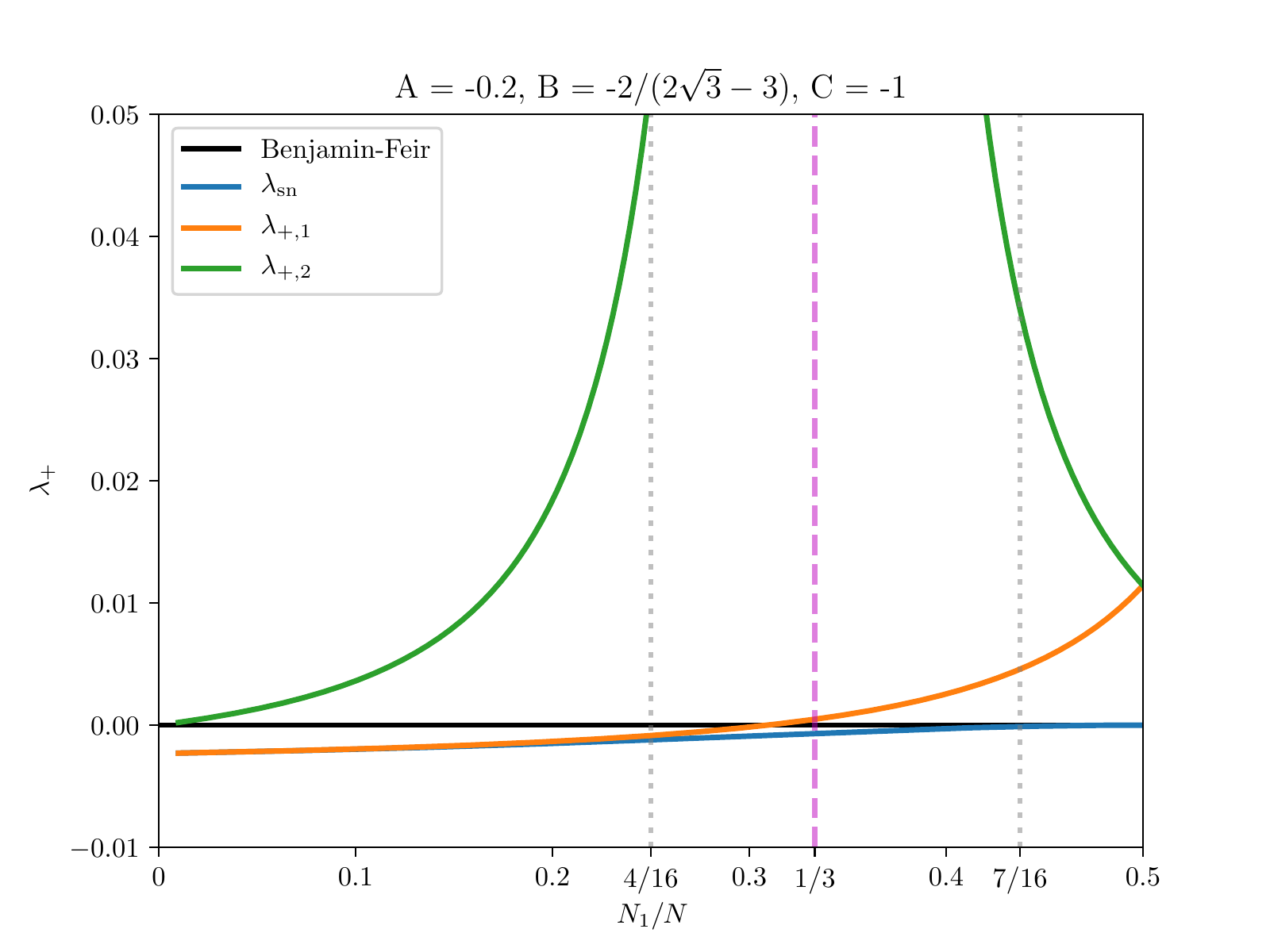}
  \caption{The bifurcation curves $\lambda_{\mathrm{sn}}$ (blue), $\lambda_{+,1}$ (orange)
    and $\lambda_{+,2}$ (green)
    for the different cluster-size distributions in the $\lambda_+$, $N_1/N$ plane
    with $A=-0.2$ and the parameters $B=-2/(2\sqrt{3}-3)$ and $C=-1$.
    The Benjamin-Feir instability is indicated by the black solid line.
    The dashed magenta line indicates the location where $\lambda_{+,2}$ diverges.
    The positions of the $4 : 12$ and $7 : 9$ cluster states are marked by the dotted vertical gray lines,
    see also Fig.~\ref{fig:sol79} for the respective solution curves.}
  \label{fig:eckhaus}
\end{figure}

The bifurcation scenario can be better visualized by plotting $\lambda_{\text{sn}}$,
$\lambda_{+,1}$ and $\lambda_{+,2}$ as a function of the cluster size $N_1/N$,
see Fig.~\ref{fig:eckhaus}.
It depicts the $\lambda_+$ values of the saddle-node bifurcations creating the
2-cluster states ($\lambda_{\text{sn}}$, blue) and of the two transverse bifurcations
(Eqs. \eqref{eq:lambda_1} and \eqref{eq:lambda_2}) altering
the stability of the 2-clusters, with $\lambda_{+,1}$ in green
and $\lambda_{+,2}$ in orange.

\begin{figure}[ht]
  \centering
  \includegraphics{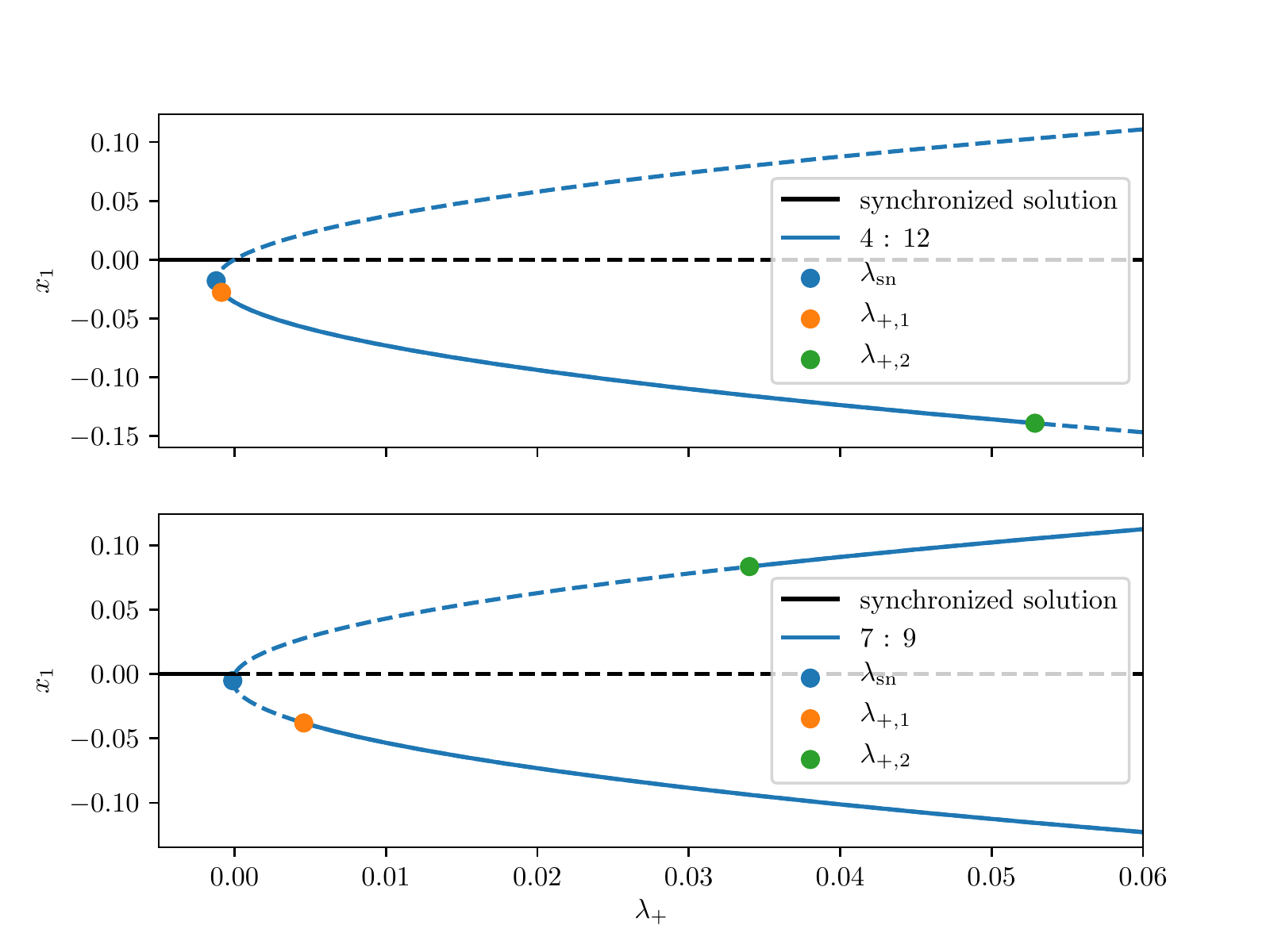}
  \caption{The variable $x_1$ of the $4 : 12$ cluster solution (top) and the $7 : 9$ cluster solution
    (bottom) as a function
    of the bifurcation parameter $\lambda_+$ with $A=-0.2$,
    and the parameters $B=-2/(2\sqrt{3}-3)$ and $C=-1$.
    Solid curves indicate that the solution is stable for the respective range of parameters,
    dashed curves represent unstable solutions.
    The points mark the $\lambda_{\mathrm{sn}}$ (blue), the $\lambda_{+,1}$ (orange)
    and the $\lambda_{+,2}$ (green) bifurcations.
    The synchronized solution $x_i=0 \, \forall i$ is indicated by the black horizontal line.
    See also Fig.~\ref{fig:eckhaus} for the locations of the $4:12$ and $7 : 9$ cluster in the $\lambda_+$,
    $N_1/N$ plane}
  \label{fig:sol79}
\end{figure}

When increasing $\lambda_+$ coming from negative values, all cluster states with $N_1/N\neq 1/2$
are born in the saddle-node bifurcation $\lambda_{\text{sn}}$.
Note that in fact two solutions for each $N_1/N$ are created this way.
In Fig.~\ref{fig:eckhaus}, one can observe that for the most unbalanced state $N_1/N\rightarrow 0$,
the transverse bifurcation stabilizing the smaller cluster $\lambda_{+,1}$ occurs immediately
after the saddle-node bifurcation creating that cluster.
This bifurcation alters the stability of one of the two solutions born in the saddle-node bifurcation,
and in particular
renders the smaller of the two clusters in that solution stable to transverse perturbations.
% alters the transverse instability of the smaller of the two clusters in that solution.
For the parameter regime considered here ($A=-0.2$, $B=-2/(2\sqrt{3}-3)$ and $C=-1$),
this solution is in fact stabilized at this bifurcation, that is for $\lambda_+>\lambda_{+,1}$.\\
For $N_1/N<1/3$, the respective 2-cluster solution remains stable until $\lambda_{+,2}$, where
the larger cluster becomes unstable, thus rendering the whole solution unstable.
This can, for example, be observed for the $4 : 12$ cluster-size distribution, see Fig.~\ref{fig:sol79}(top).
There, the variable of one cluster, $x_1$, is plotted as a function of the bifurcation parameter
$\lambda_+$.
The blue dot on the left marks the saddle-node bifurcation wherein the two $4:12$ solutions are created.
Initially, both solutions are unstable.
At $\lambda_{+,1}$ (orange dot), one of them is stabilized, and at $\lambda_{+,2}$ (green dot), it is subsequently destabilized.\\
%% The saddle node creating the solutions of this 2-cluster state is marked by a blue dot, where the
%% two solution curves emerge.
%% Note that one of the two solutions becomes stabilized at $\lambda_{+,1}$ (orange dot) and subsequently
%% destabilized at $\lambda_{+,2}$ (green dot).
For $N_1/N>1/3$, the scenario is different.
There the solution that got stabilized at $\lambda_{+,1}$ remains stable for all $\lambda_+>\lambda_{+,1}$.
The bifurcation $\lambda_{+,2}$ instead occurs at the second cluster solution created at
the saddle-node bifurcation.
This is illustrated more clearly in Fig.~\ref{fig:sol79}(bottom) for the $7 : 9$ cluster solution.
One of the two solutions becomes stable at $\lambda_{+, 1}$, marked by an
orange dot and as discussed above. Since $N_1/N = 7/16>1/3$, this solution remains
stable for all $\lambda_+>\lambda_{+, 1}$.
The second solution (upper curve in the bottom part of Fig.~\ref{fig:sol79}) first passes the synchronized solution
at the Benjamin-Feir bifurcation $\lambda_+=0$ and finally becomes stabilized at $\lambda_{+, 2}$,
marked by a green dot.
$\lambda_{+, 2}$ diverges at the pole $N_1/N=1/3$, separating the two scenarios shown in
Fig.~\ref{fig:sol79}.
There the bifurcation switches from the solution with negative $x_1$ (which, for $\lambda_+ \rightarrow \infty$, diverges to $-\infty$) to the solution with positive $x_1$ (which, for $\lambda_+ \rightarrow \infty$, diverges to $+\infty$).\\
Notice how for the cluster distribution $N_1/N=7/16$ the two 2-cluster solutions are bistable for
$\lambda_+>\lambda_{+, 2}$. That is, there exist two stable 2-cluster solutions with different $x_1$
but the same cluster size ratio $7 : 9$ that are both stable.
This, in fact, has also been observed in the Stuart-Landau ensemble, see for example Fig. 6 in
Ref.~\cite{kemeth19_clust_singul}.
Note that the singularity of $\lambda_{+, 2}$ at $N_1/N=1/3$ ($\alpha=1/2$)
is independent of the parameters
$A$, $B$ and $C$, see Eq.~\eqref{eq:lambda_2}.
This means that bistable solutions created as described above
can in general only exist for $N_1/N>1/3$.

\section{Conclusion and Outlook}
\label{sec:conclusion}

In this paper, we showed how one can map a system of globally coupled Stuart-Landau oscillators onto the
$(N-1)$-dimensional center manifold at the Benjamin-Feir instability.
Thereby, we observed that the bifurcation curves at which 2-cluster solutions are born closely resemble
their counterparts in the original oscillatory system.
This allowed us to investigate a codimension-two point called cluster singularity,
from which all these bifurcation curves emanate.
In the center manifold, we saw that this point corresponds to a vanishing coefficient $A=0$ in front
of the quadratic term of the equations of motion.
Due to the reduced dynamics in this manifold, we were able
to obtain stability boundaries for 2-cluster states analytically.
This allows for the more detailed investigation of the bow-tie-shaped cascade of
transverse bifurcations that govern the stability of these 2-cluster states,
highlighting the role of the cluster singularity as an organizing center.
%% this allows for the investigation of the bow-tie-shaped cascade of
%% transverse bifurcations in more detail,
%% highlighting the role of the cluster singularity as an organizing center.
The observed behavior is hereby independent of the
oscillatory nature of each Stuart-Landau oscillator,
but a result of the $\mathbf{S}_N$-equivariance of the full system.
These findings may thus facilitate our understanding of this codimension-two point,
and of clustering in general, even beyond oscillatory ensembles.\\
Through this reduction to the center manifold, we could calculate the bifurcation curves creating
the cluster solutions ($\lambda_{\mathrm{sn}}$) and altering their stability
($\lambda_{+, 1}$ and $\lambda_{+, 2}$) analytically.
This allowed us to investigate when stable 2-cluster solutions exist more
systematically, and in particular revealed when
different solutions with the same cluster-size distribution are bistable (cf. Fig.~\ref{fig:sol79}).
%phenomena such as bistability of
% particular cluster distributions  can occur.
The relative cluster size $N_1/N=1/3$ seems to be a general lower
limit for such a bistable behavior.
%% Furthermore, the role of this codimension-two point can be understood as an
%% organizing center: All bifurcation curves creating 2-cluster states,
%% as well as secondary bifurcation curves altering their stability, emanate from this point.
The bifurcation scenario of how states with different cluster
size ratios $N_1/N$ are created is thereby different from the
Eckhaus instability~\cite{tuckerman90_bifur_analy_eckhaus_instab} in reaction-diffusion systems.
There, solutions of different wavelengths are created through
supercritial pitchfork bifurcations at the trivial solution and subsequently
stabilized through a sequence of subcritical pitchfork bifurcations involving mixed-mode
states.
In our case, the different 2-cluster states are created in saddle-node bifurcations
and stabilized at $\lambda_{+, 1}$ at a single equivariant bifurcation point
involving 3-cluster states.
However, the detailed interaction between 2- and 3-cluster states still remains
an open topic for future research.\\
Note that the cubic truncation of the flow in the center manifold has a
gradient structure~\cite{fiedler20_dynamics}.
This means that we can assign an abstract potential to each of the cluster distributions for a
particular set of parameters $\lambda_+, A, B$ and $C$.
Is there a particular cluster distribution with a minimal potential value?
What is its role in the dynamics between these cluster distributions?
The companion paper~\cite{fiedler20_dynamics} addresses some of these dynamical questions.\\
Here, we fixed the parameter $\gamma=2$ in the full Stuart-Landau system,
and varied the coupling parameters $\beta_{\textrm{r}}$, $\beta_{\textrm{i}}$.
This restricts our analysis to a small region in parameter space.
It is important to mention that for different parameter regimes, a qualitatively different
behavior close to the cluster singularity might be observed~\cite{fiedler20_dynamics}.\\
As discussed in Sec.~\ref{sec:var_trans},
the Stuart-Landau ensemble permits the transformation into a corotating frame.
This turns limit-cycle dynamics into fixed-point dynamics and thus greatly
facilitates the reduction onto the center manifold.
For more general oscillatory ensembles, such as systems composed of
van der Pol or Hogdkin-Huxley type units, the transformation
to a corotating frame may be more cumbersome or not even possible.
If the coupling between such units is of a global nature,
we expect, however, that the nesting of bifurcation curves creating different cluster
distributions,
cf. Fig.~\ref{fig:sn_curves_cmf}, can also be observed in these systems.\\
%% Is there an analogous way to reduce these systems to the center manifold?\\
This directly links to the fact that we focused on oscillatory dynamics in this article.
An exciting further question is the possibility of equivalent dynamics, such as clustering
and cluster singularities, in systems composed of bistable or excitable units.

\section*{Acknowledgement}

FPK thanks BF for the
hospitality and the exciting discussions at the Freie Universität Berlin.
BF gratefully acknowledges the deep inspiration by,
and hospitality of, his coauthors in München who initiated this work.
This work has also been supported by the Deutsche Forschungsgemeinschaft,
SFB910, project A4 ``Spatio-Temporal Patterns: Control, Delays, and Design'',
and by KR1189/18 ``Chimera States and Beyond''.

%% We parametrize this center manifold using the coordinates $x_k$ and $y_k$ with
%% \begin{align*}
%% 2dx_k & = -r_k + \frac{d+1}
%% \end{align*}
\appendix
\section{Variable transformation}
\label{sec:aaa}

Using log-polar coordinates $W_k=\exp\left(R_k+i\Phi_k\right)$, Eq.~\eqref{eq:sle} turns into
\begin{equation*}
  \left(\dot{R}_k + i\dot{\Phi}_k\right)e^{R_k+i\Phi_k}
  = e^{R_k+i\Phi_k}-\left(1+i\gamma\right)e^{2R_k}e^{R_k+i\Phi_k}+
 \left(\beta_{\textrm{r}} + i\beta_{\textrm{i}}\right)\left(\langle e^{\mathbf{R}+i\mathbf{\Phi}}\rangle-e^{R_k+i\Phi_k}\right).
\end{equation*}
Dividing by $W_k$ this becomes
\begin{equation*}
  \dot{R}_k + i\dot{\Phi}_k
  = 1-\left(1+i\gamma\right)e^{2R_k}+
  \left(\beta_{\textrm{r}} + i\beta_{\textrm{i}}\right)
  \left(\langle e^{\mathbf{R}+i\mathbf{\Phi}}\rangle e^{-R_k-i\Phi_k}-1\right).
\end{equation*}
We average over $k$ and separate real and imaginary parts.
The mean amplitude $R$ and the mean phase $\Phi$ then satisfy
\begin{align*}
  \dot{R} & = 1-\langle e^{2\mathbf{R}} \rangle+
  \operatorname{Re}\left(\left(\beta_{\textrm{r}} + i\beta_{\textrm{i}}\right)\left(\langle e^{\mathbf{R}+i\mathbf{\Phi}}\rangle \langle e^{-\mathbf{R}-i\mathbf{\Phi}}\rangle-1\right)\right)\\
  \dot{\Phi} & = -\gamma\langle e^{2\mathbf{R}} \rangle+
  \operatorname{Im}\left(\left(\beta_{\textrm{r}} + i\beta_{\textrm{i}}\right)\left(\langle e^{\mathbf{R}+i\mathbf{\Phi}}\rangle \langle e^{-\mathbf{R}-i\mathbf{\Phi}}\rangle-1\right)\right).
\end{align*}
Substituting the variables listed in Tab.~\ref{tab:co_trans},
one obtains $\langle \exp\left(2\mathbf{R}\right) \rangle=\langle \exp\left(2\mathbf{r}+2R\right) \rangle
=\exp\left(2R\right)\langle \exp\left(2\mathbf{r}\right)\rangle$,
and $\langle \exp\left(\mathbf{R}+i\mathbf{\Phi}\right)\rangle = \langle \exp\left(\mathbf{r}+R+i\boldsymbol{\varphi}+i\Phi\right)\rangle=\exp\left(R+i\Phi\right)\langle \exp \mathbf{z}\rangle$.
Therefore
\begin{align*}
  \dot{R} & = 1-e^{2R}\langle e^{2\mathbf{r}} \rangle+
  \operatorname{Re}\left(\left(\beta_{\textrm{r}} + i\beta_{\textrm{i}}\right)\left(\langle e^{\mathbf{z}}\rangle \langle e^{-\mathbf{z}}\rangle-1\right)\right)\\
  \dot{\Phi} & = -\gamma e^{2R}\langle e^{2\mathbf{r}} \rangle+
  \operatorname{Im}\left(\left(\beta_{\textrm{r}} + i\beta_{\textrm{i}}\right)\left(\langle e^\mathbf{z}\rangle \langle e^{-\mathbf{z}}\rangle-1\right)\right).
\end{align*}
For the deviations $r_k = R_k-R$ and $\varphi_k = \Phi_k-\Phi$ one may write
\begin{align*}
  \dot{r}_k & = \dot{R}_k -\dot{R}\\
  & = 1-e^{2R}e^{2r_k}+\operatorname{Re}\left(\left(\beta_{\textrm{r}} + i\beta_{\textrm{i}}\right)\left(\langle e^{\mathbf{z}}\rangle e^{-z_k} -1\right)\right)-\dot{R}\\
  & = -e^{2R}\widetilde{e^{2r_k}}+\operatorname{Re}\left(\left(\beta_{\textrm{r}} + i\beta_{\textrm{i}}\right)\left(\langle e^{\mathbf{z}}\rangle\widetilde{ e^{-z_k}}\right)\right)\\
  \dot{\varphi}_k & = \dot{\Phi}_k -\dot{\Phi}\\
  & = -\gamma e^{2R}e^{2r_k}+\operatorname{Im}\left(\left(\beta_{\textrm{r}} + i\beta_{\textrm{i}}\right)\left(\langle e^{\mathbf{z}}\rangle e^{-z_k} -1\right)\right)-\dot{\Phi}\\
    & = -\gamma e^{2R}\widetilde{e^{2r_k}}+\operatorname{Im}\left(\left(\beta_{\textrm{r}} + i\beta_{\textrm{i}}\right)\left(\langle e^{\mathbf{z}}\rangle \widetilde{e^{-z_k}}\right)\right)
\end{align*}
with the notations defined in table~\ref{tab:nots}. The equations for $\dot{R}$, $\dot{r}_k$ and $\dot{\varphi}_k$ then constitute the corotating system Eqs.~\eqref{eq:3a} to~\eqref{eq:3c}.

\section{Linearization}
\label{sec:aa}
Linearizing the dynamics of the transformed system, Eqs.~\eqref{eq:3a} to~\eqref{eq:3c},
at the equilibrium $R=0$, $r_k=\varphi_k=0$, $z_k=0$, 
and using the fact that $\langle \mathbf{r}\rangle=0$, $\langle \mathbf{z}\rangle=0$, see Tab.~\ref{tab:co_trans},
one gets
\begin{align*}
  \begin{pmatrix}
    \dot{R}\\ \dot{r}_k \\ \dot{\varphi}_k
  \end{pmatrix} & =\begin{pmatrix}
  -2R\\
  -2r_k - \operatorname{Re}\left(kz_k\right)\\
  -2\gamma r_k- \operatorname{Im}\left(kz_k\right)
  \end{pmatrix}\\
  & = \begin{pmatrix}
  -2R\\
  -\left(2+\beta_{\textrm{r}}\right)r_k+\beta_{\textrm{i}}\varphi_k\\
  -\left(2\gamma+\beta_{\textrm{i}}\right)r_k-\beta_{\textrm{r}}\varphi_k
  \end{pmatrix}\\
  & = \begin{pmatrix}
  -2 & 0 & 0 \\
  0 & -2-\beta_{\textrm{r}} & \beta_{\textrm{i}}\\
  0 & -2\gamma-\beta_{\textrm{i}} & -\beta_{\textrm{r}}
  \end{pmatrix} \cdot \begin{pmatrix}
    R\\ r_k \\ \varphi_k
  \end{pmatrix}=\mathbf{J}\cdot \begin{pmatrix}
    R\\ r_k \\ \varphi_k
  \end{pmatrix}.
\end{align*}
The Jacobian thus has the eigenvalues
\begin{itemize}
\item Eigenvalue $\lambda_1 = -2$ with eigenvector $\vec{v}_1=\left(1, \vec{0}, \vec{0}\right)$.
\end{itemize}
and two eigenvalues of geometric multiplicity $N-1$ given by the eigendecomposition
\begin{equation*}
\text{eig}\begin{pmatrix}[1.75]
   -2-\beta_{\textrm{r}} & \beta_{\textrm{i}}\\
   -2\gamma-\beta_{\textrm{i}} & -\beta_{\textrm{r}}
  \end{pmatrix},
\end{equation*}
which gives
\begin{itemize}
\item the eigenvalue $\lambda_+ = -1-\beta_{\textrm{r}}+\sqrt{1-\beta_{\textrm{i}}^2-2\beta_{\textrm{i}}\gamma} = -1-\beta_{\textrm{r}}+d$ % with eigenvector $\vec{v}_+=\left(0, x_k, (1+d)/\beta_{\textrm{i}} x_k\right)$.
\item and the eigenvalue $\lambda_- = -1-\beta_{\textrm{r}}-\sqrt{1-\beta_{\textrm{i}}^2-2\beta_{\textrm{i}}\gamma} = -1-\beta_{\textrm{r}}-d$. %with eigenvector $\vec{v}_-=\left(0, x_k, (1-d)/\beta_{\textrm{i}} x_k\right)$.
\end{itemize}
Here, we assume $1-\beta_{\textrm{i}}^2-2\beta_{\textrm{i}}\gamma>0$, that is real $\lambda_\pm$. For an analysis of the case
$1-\beta_{\textrm{i}}^2-2\beta_{\textrm{i}}\gamma<0$, see Ref.~\cite{fpkemeth19_thesis}.
The eigenvectors corresponding to these two eigenvalues can be obtained using
\begin{equation*}
\left(\begin{pmatrix}[1.75]
   -2-\beta_{\textrm{r}} & \beta_{\textrm{i}}\\
   -2\gamma-\beta_{\textrm{i}} & -\beta_{\textrm{r}}
  \end{pmatrix}
-\lambda_{\pm}\mathbf{1}_{(N-1)\times (N-1)}\right) \vec{v}_\pm = \vec{0}.
\end{equation*}
For $\lambda_+$, one thus obtains
\begin{align*}
  \begin{pmatrix}
-1-d & \beta_{\textrm{i}}\\
-2\gamma-\beta_{\textrm{i}} & 1-d
  \end{pmatrix}\vec{v}_+ & =   \begin{pmatrix}
-1-d & \beta_{\textrm{i}}\\
-2\gamma-\beta_{\textrm{i}} & 1-d
  \end{pmatrix}\begin{pmatrix}
    r_k\\
    \varphi_k
  \end{pmatrix} =   \begin{pmatrix}
\left(-1-d\right)r_k + \beta_{\textrm{i}}\varphi_k\\
\left(-2\gamma-\beta_{\textrm{i}}\right)r_k + \left(1-d\right)\varphi_k
  \end{pmatrix}=0.
\end{align*}
Choosing
\begin{equation}
  \varphi_k = (1+d)/\beta_{\textrm{i}} r_k,\tag{A.1}\label{eq:constp}
\end{equation}
we get
\begin{align*}
  \begin{pmatrix}
\left(-1-d\right)r_k+ \left(1+d\right)r_k\\
\left(-2\gamma-\beta_{\textrm{i}}\right)r_k + \left(1-d^2\right)/\beta_{\textrm{i}} r_k
  \end{pmatrix}=  \begin{pmatrix}
-\left(1+d\right)r_k+ \left(1+d\right)r_k\\
-\left(2\gamma+\beta_{\textrm{i}}\right)r_k + \left(2\gamma+\beta_{\textrm{i}}\right) r_k
  \end{pmatrix} = \vec{0},
\end{align*}
thus solving the equality above.
The constraint Eq.~\eqref{eq:constp}, together with $\langle \mathbf{r}\rangle=\langle\boldsymbol{\varphi}\rangle=0$, defines an $(N-1)$-dimensional subspace of $\mathbb{R}^{2N-1}$.\\
For $\lambda_-$, one thus obtains
\begin{align*}
  \begin{pmatrix}
-1+d & \beta_{\textrm{i}}\\
-2\gamma-\beta_{\textrm{i}} & 1+d
  \end{pmatrix}\vec{v}_+ & =   \begin{pmatrix}
-1+d & \beta_{\textrm{i}}\\
-2\gamma-\beta_{\textrm{i}} & 1+d
  \end{pmatrix}\begin{pmatrix}
    r_k\\
    \varphi_k
  \end{pmatrix} =   \begin{pmatrix}
\left(-1+d\right)r_k & \beta_{\textrm{i}}\varphi_k\\
\left(-2\gamma-\beta_{\textrm{i}}\right)r_k & \left(1+d\right)\varphi_k
  \end{pmatrix}.
\end{align*}
Choosing
\begin{equation}
  \varphi_k = (1-d)/\beta_{\textrm{i}} r_k,\tag{A.2}\label{eq:constm}
\end{equation}
solves the conditions above. In particular,
\begin{align*}
  \begin{pmatrix}
\left(-1+d\right)r_k+ \left(1-d\right)r_k\\
\left(-2\gamma-\beta_{\textrm{i}}\right)r_k + \left(1-d^2\right)/\beta_{\textrm{i}} r_k
  \end{pmatrix}=  \begin{pmatrix}
-\left(1-d\right)r_k+ \left(1-d\right)r_k\\
-\left(2\gamma+\beta_{\textrm{i}}\right)r_k + \left(2\gamma+\beta_{\textrm{i}}\right) r_k
  \end{pmatrix} = \vec{0}.
\end{align*}
The constraint Eq.~\eqref{eq:constm}, together with $\langle \mathbf{r}\rangle=\langle\boldsymbol{\varphi}\rangle=0$ define an $(N-1)$-dimensional subspace of $\mathbb{R}^{2N-1}$.\\
Now, one can define the eigencoordinates $x_k$ describing the dynamics in the space defined by the constraint Eq.~\eqref{eq:constp}, the center space of the bifurcation, and eigencoordinates $y_k$,
describing the dynamics in the space defined by the constraint Eq.~\eqref{eq:constm}.
These two sets of variables, together with $R$, can then be used to describe the full system.

\section{Parameter Derivation}
\label{sec:par_ests}
In this section of the appendix, we derive expressions for the parameters
$a$, $b$, $A$, $B$ and $C$ as a function of the parameters $\gamma$, $\beta_{\textrm{r}}$ and $\beta_{\textrm{i}}$ from
the Stuart-Landau ensemble.
Hereby, we will use the condition that $R$ and the $y_k$ are tangential,
that is, $\left. \frac{\mathrm{d}}{\mathrm{d}x_k} R\right|_{\mathbf{x}=0}=0$ and
$\left. \frac{\mathrm{d}}{\mathrm{d}x_k} y_k\right|_{\mathbf{x}=0}=0$.
\subsection{$a$ and $b$}
In order to calculate $a$ and $b$, it is useful to write out the following expressions 
\begin{align*}
  z_k & = r_k+i\varphi_k\\
  & = \left(1-d\right) x_k + \left(1+d\right)y_k+ i\left(\gamma' x_k + \gamma' y_k\right)\\
  & = \left(1-d+i\gamma'\right)x_k+a\left(1+d+i\gamma'\right)\widetilde{x_k^2}
  + \mathcal{O}\left(x_k^3\right)\\
  z_k^2 & = \left(r_k+i\varphi_k\right)^2\\
  & = \left(\left(1-d\right) x_k + \left(1+d\right)y_k+ i\left(\gamma' x_k + \gamma' y_k\right)\right)^2\\
  & = \left(\left(1-d+i\gamma'\right) x_k + \left(1+d+i\gamma'\right)y_k\right)^2\\
  & = \left(1-d+i\gamma'\right)^2 x_k^2 +
  2a\left(1-d+i\gamma'\right)\left(1+d+i\gamma'\right)x_k\widetilde{x_k^2}+ \mathcal{O}\left(x_k^4\right)\\
  z_k^3  & = \left(r_k+i\varphi_k\right)^3\\
  & = \left(\left(1-d+i\gamma'\right) x_k + \left(1+d+i\gamma'\right)y_k\right)^3\\
  & = \left(1-d+i\gamma'\right)^3x_k^3  + \mathcal{O}\left(x_k^4\right)
\end{align*}
where we used Eq~\eqref{eq:yk} for $y_k$ and the notation $\gamma' = 2\gamma+\beta_{\textrm{i}}$.
Similarly, we expand the following parts and keep terms up to cubic order:
\begin{align*}
  e^{z_k} & = 1+z_k+\frac{z_k^2}{2} + \frac{z_k^3}{6} + \mathcal{O}\left(x_k^4\right)\\
  e^{-z_k} & = 1-z_k+\frac{z_k^2}{2} - \frac{z_k^3}{6} + \mathcal{O}\left(x_k^4\right)\\
  \langle e^{\mathbf{z}}\rangle & = \langle 1+\mathbf{z} + \frac{\mathbf{z}^2}{2} +
  \frac{\mathbf{z}^3}{6} +
  \mathcal{O}\left(x_k^4\right)\rangle\\
  & = 1 + \frac{1}{2} \langle \mathbf{z}^2\rangle + \frac{1}{6}\langle{\mathbf{z}^3}\rangle +
  \mathcal{O}\left(x_k^4\right)\\
  \widetilde{e^{-z_k}} &= e^{-z_k}-\langle e^{-\mathbf{z}}\rangle\\
  & = 1-z_k+\frac{z_k^2}{2} - \frac{z_k^3}{6} - 1 -\frac{1}{2}\langle \mathbf{z}^2\rangle
  + \frac{1}{6}\langle \mathbf{z}^3 \rangle +\mathcal{O}\left(x_k^4\right)\\
  & = -z_k + \frac{1}{2}\widetilde{z_k^2} - \frac{1}{6}\widetilde{z_k^3}
  + \mathcal{O}\left(x_k^4\right)\\
\end{align*}
\begin{align*}
  \langle e^{\mathbf{z}}\rangle \langle e^{-\mathbf{z}}\rangle & =
  \left(1 + \frac{1}{2} \langle \mathbf{z}^2\rangle + \frac{1}{6}\langle{\mathbf{z}^3}\rangle\right)
  \left(1 + \frac{1}{2} \langle \mathbf{z}^2\rangle - \frac{1}{6}\langle{\mathbf{z}^3}\rangle \right)
  + \mathcal{O}\left(x_k^4\right)\\
  & = 1 + \frac{1}{2} \langle \mathbf{z}^2\rangle + \frac{1}{6}\langle{\mathbf{z}^3}\rangle +
  \frac{1}{2} \langle \mathbf{z}^2\rangle - \frac{1}{6}\langle{\mathbf{z}^3}\rangle
  + \mathcal{O}\left(x_k^4\right)\\
  & = 1 + \langle \mathbf{z}^2\rangle + \mathcal{O}\left(x_k^4\right)\\
  \langle e^{\mathbf{z}}\rangle \widetilde{e^{-z_k}} & =
  \left(1 + \frac{1}{2} \langle \mathbf{z}^2\rangle + \frac{1}{6}\langle{\mathbf{z}^3}\rangle\right)
  \left(-z_k + \frac{1}{2}\widetilde{z_k^2} - \frac{1}{6}\widetilde{z_k^3}\right)
  + \mathcal{O}\left(x_k^4\right)\\
  & = \left(1 + \frac{1}{2} \langle \mathbf{z}^2\rangle\right)
  \left(-z_k + \frac{1}{2}\widetilde{z_k^2} - \frac{1}{6}\widetilde{z_k^3}\right)
  + \mathcal{O}\left(x_k^4\right)\\
  & = -z_k + \frac{1}{2}\widetilde{z_k^2} - \frac{1}{6}\widetilde{z_k^3}
  -\frac{1}{2} z_k \langle \mathbf{z}^2\rangle  + \mathcal{O}\left(x_k^4\right).\\
\end{align*}
With the expression for $R$, see Eq.~\eqref{eq:R}, we can furthermore write
\begin{align*}
  e^{2R} & = 1+2R + \mathcal{O}\left(x_k^4\right)\\
  & = 1+2b\langle \mathbf{x}^2\rangle+ \mathcal{O}\left(x_k^4\right)\\
  e^{2r_k} & = 1+2r_k+2r_k^2+\frac{4}{3} r_k^3+ \mathcal{O}\left(x_k^4\right)\\
  \langle e^{2\mathbf{r}} \rangle & = 1+2\langle \mathbf{r}^2\rangle +
  \frac{4}{3} \langle \mathbf{r}^3\rangle+ \mathcal{O}\left(x_k^4\right)\\
  \widetilde{e^{2r_k}} & = e^{2r_k} - \langle e^{2\mathbf{r}} \rangle\\
  & = 2r_k + 2\widetilde{r_k^2}+\frac{4}{3}\widetilde{r_k^3}+ \mathcal{O}\left(x_k^4\right)\\
  e^{2R}\langle e^{2\mathbf{r}} \rangle & = \left(1+2b\langle \mathbf{x}^2\rangle \right)
  \left(1+2\langle \mathbf{r}^2\rangle +
  \frac{4}{3} \langle \mathbf{r}^3\rangle\right)+ \mathcal{O}\left(x_k^4\right)\\
  & = 1+2\langle \mathbf{r}^2\rangle +2b\langle \mathbf{x}^2\rangle + \frac{4}{3} \langle \mathbf{r}^3\rangle
  + \mathcal{O}\left(x_k^4\right)\\
  e^{2R}\widetilde{e^{2r_k}} & = \left(1+2b\langle \mathbf{x}^2\rangle\right)
  \left(2r_k + 2\widetilde{r_k^2}+\frac{4}{3}\widetilde{r_k^3}\right)+ \mathcal{O}\left(x_k^4\right)\\
  & = 2r_k + 2br_k\langle \mathbf{x}^2\rangle + 2\widetilde{r_k^2}+\frac{4}{3}\widetilde{r_k^3}
  + \mathcal{O}\left(x_k^4\right).
\end{align*}
Using these approximations, we can write for the dynamics of $R$ up to second order in $x_k$
\begin{align*}
\dot{R} & = 1-e^{2R} \langle e^{2\mathbf{r}}\rangle +
\operatorname{Re} \left(\left(\beta_{\textrm{r}} + i\beta_{\textrm{i}}\right)\left(\langle e^{\mathbf{z}}\rangle \langle e^{-\mathbf{z}} \rangle-1\right)\right)\\
& = 1-\left(1+2\langle \mathbf{r}^2\rangle +2b\langle \mathbf{x}^2\rangle\right)
+ \operatorname{Re}\left(\left(\beta_{\textrm{r}} + i\beta_{\textrm{i}}\right) \left(1 + \langle \mathbf{z}^2\rangle-1\right)\right)\\
& = -2\langle \mathbf{r}^2\rangle -2b\langle \mathbf{x}^2\rangle
+ \operatorname{Re}\left(\left(\beta_{\textrm{r}} + i\beta_{\textrm{i}}\right) \langle \mathbf{z}^2\rangle\right)\\
& = -2\left(1-d\right)^2\langle \mathbf{x}^2\rangle
-2b\langle \mathbf{x}^2\rangle
+\operatorname{Re}\left(\left(\beta_{\textrm{r}} + i\beta_{\textrm{i}}\right) \left(1-d+i\gamma'\right)^2\right) \langle \mathbf{x}^2\rangle\\
& = -2\left(1-d\right)^2\langle \mathbf{x}^2\rangle
-2b\langle \mathbf{x}^2\rangle
+\left(\beta_{\textrm{r}} \left(\left(1-d\right)^2-\gamma'^2\right)-2\beta_{\textrm{i}}\left(\gamma'\left(1-d\right)\right)\right) \langle \mathbf{x}^2\rangle\\
& = -2\beta_{\textrm{r}}^2\langle \mathbf{x}^2\rangle
-2b\langle \mathbf{x}^2\rangle
+\left(\beta_{\textrm{r}} \left(\beta_{\textrm{r}}^2-\gamma'^2\right)-
2\left(\beta_{\textrm{r}}^2+2\beta_{\textrm{r}}\right)\beta_{\textrm{r}}\right) \langle \mathbf{x}^2\rangle\\
& = -\left(2\beta_{\textrm{r}}^2-\beta_{\textrm{r}} \left(\beta_{\textrm{r}}^2-\gamma'^2\right)+
2\left(\beta_{\textrm{r}}^2+2\beta_{\textrm{r}}\right)\beta_{\textrm{r}}{-}2b\right)\langle \mathbf{x}^2\rangle\\
& = -\left(6\beta_{\textrm{r}}^2+\beta_{\textrm{r}}^3+\beta_{\textrm{r}} \gamma'^2{+}2b\right)\langle \mathbf{x}^2\rangle
\end{align*}
Now, we use the tangential property of $R$. In particular, we can write
\begin{equation*}
  \dot{R} = \left(\frac{\mathrm{d}}{\mathrm{d}x_k} R\right) \, \dot{\mathbf{x}} =
  2b \langle \mathbf{x}\dot{\mathbf{x}}\rangle +\mathcal{O}\left(x_k^5\right) =
  2b\lambda_+ \langle \mathbf{x}^2 \rangle+
  \mathcal{O}\left(x_k^3\right).
\end{equation*}
At $\lambda_+=0$, $\dot{R}$ up to second order must vanish. This allows us to calculate $b$ by comparing the terms
in front of $\langle \mathbf{x}^2\rangle$ in $\dot{R}$, yielding
\begin{align*}
\Rightarrow b & = -\frac{\beta_{\textrm{r}}}{2}\left(\gamma'^2 +6\beta_{\textrm{r}}+\beta_{\textrm{r}}^2\right)\\
& = \frac{1-d}{2}\left(\gamma'^2 +d^2+4d-5\right).
\end{align*}
We can derive the expression for $a$ in a similar way.
Here, we write out the dynamics of $y_k$ up to second order.
This yields
\begin{align*}
  2d\dot{y}_k & = \dot{r}_k + \frac{d-1}{\gamma'} \dot{\varphi}_k\\
  & = -\left(1+\left(d-1\right)\frac{\gamma}{\gamma'}\right)e^{2R}\widetilde{e^{2r_k}} +
  \operatorname{Re}\left(\left(1-i\frac{d-1}{\gamma'}\right)\left(\beta_{\textrm{r}} + i\beta_{\textrm{i}}\right)
  \left(\langle e^{\mathbf{z}}\rangle \widetilde{e^{-z_k}}\right)\right)\\
  & = -\left(1+\left(d-1\right)\frac{\gamma}{\gamma'}\right)
  \left(2r_k + 2\widetilde{r_k^2}\right)
  + \operatorname{Re}\left(\left(1-i\frac{d-1}{\gamma'}\right)\left(\beta_{\textrm{r}} + i\beta_{\textrm{i}}\right)
  \left(-z_k + \frac{1}{2}\widetilde{z_k^2}\right)\right)\\
  %% & = -\left(1+\left(d-1\right)\frac{\gamma}{\gamma'}\right)
  %% \left(2r_k + 2\widetilde{r_k^2}\right)
  %% + \operatorname{Re}\left(\left(1-i\frac{d-1}{\gamma'}\right)\left(\beta_{\textrm{r}} + i\beta_{\textrm{i}}\right)
  %% \left(-z_k + \frac{1}{2}\widetilde{z_k^2}\right)\right)\\
  & = -\left(1+\left(d-1\right)\frac{\gamma}{\gamma'}\right)
  \left(2r_k + 2\widetilde{r_k^2}\right)
  + \operatorname{Re}\left(\left(1-i\frac{d-1}{\gamma'}\right)\left(\beta_{\textrm{r}} + i\beta_{\textrm{i}}\right)
  \left(-r_k-i\varphi_k + \frac{1}{2}  {\widetilde{z_k^2}}\right)\right).
\end{align*}
The term of the coupling constant and its parameters in front can be summarized by
\begin{align*}
  \left(1-i\frac{\beta_{\textrm{r}}}{\gamma'}\right)\left(\beta_{\textrm{r}}+i\beta_{\textrm{i}}\right) & =
  \beta_{\textrm{r}}+\frac{\beta_{\textrm{i}}\beta_{\textrm{r}}}{\gamma'}-i\left(\frac{\beta_{\textrm{r}}^2}{\gamma'}-\beta_{\textrm{i}}\right)\\
  & = \beta_{\textrm{r}}-\frac{\beta_{\textrm{r}}^3+2\beta_{\textrm{r}}^2}{\gamma'^2}
  -i\left(\frac{\beta_{\textrm{r}}^2}{\gamma'}+\frac{\beta_{\textrm{r}}^2+2\beta_{\textrm{r}}}{\gamma'}\right)\\
  \beta_{\textrm{r}}\frac{\gamma}{\gamma'} & = \beta_{\textrm{r}}\frac{\gamma'-\beta_{\textrm{i}}}{2\gamma'}\\
  & =\frac{\beta_{\textrm{r}}}{2} + \frac{\beta_{\textrm{r}}^3+2\beta_{\textrm{r}}^2}{2\gamma'^2}.
\end{align*}
This simplifies the expression for $\dot{y}_k$ to
\begin{align*}
  2d\dot{y}_k   & = -\left(2+\beta_{\textrm{r}} + \frac{\beta_{\textrm{r}}^3+2\beta_{\textrm{r}}^2}{\gamma'^2}\right)
  \left(r_k + \widetilde{r_k^2}\right)\\
  & + \operatorname{Re}\left(\left(\beta_{\textrm{r}}-\frac{\beta_{\textrm{r}}^3+2\beta_{\textrm{r}}^2}{\gamma'^2}
  -i\left(\frac{\beta_{\textrm{r}}^2}{\gamma'}+\frac{\beta_{\textrm{r}}^2+2\beta_{\textrm{r}}}{\gamma'}\right)\right)
  \left(-r_k-i\varphi_k + \frac{1}{2}  {\widetilde{z_k^2}}\right)\right)\\
  & = -\left(2+\beta_{\textrm{r}} + \frac{\beta_{\textrm{r}}^3+2\beta_{\textrm{r}}^2}{\gamma'^2}\right)
  \left(r_k + \widetilde{r_k^2}\right)\\
  & + \operatorname{Re}\left(\left(\beta_{\textrm{r}}-\frac{\beta_{\textrm{r}}^3+2\beta_{\textrm{r}}^2}{\gamma'^2}
  -i\left(\frac{\beta_{\textrm{r}}^2}{\gamma'}+\frac{\beta_{\textrm{r}}^2+2\beta_{\textrm{r}}}{\gamma'}\right)\right)
  \left(-r_k-i\varphi_k +\frac{1}{2}\left(\left(1-d\right)^2-\gamma'^2\right)\widetilde{ x_k^2}+
  i\left(1-d\right)\gamma'\widetilde{ x_k^2}\right)\right)
\end{align*}
\begin{align*}
  & = -\left(2+\beta_{\textrm{r}} + \frac{\beta_{\textrm{r}}^3+2\beta_{\textrm{r}}^2}{\gamma'^2}\right)
  \left(r_k + \widetilde{r_k^2}\right)\\
  & +\left(\beta_{\textrm{r}}-\frac{\beta_{\textrm{r}}^3+2\beta_{\textrm{r}}^2}{\gamma'^2}\right)
  \left(-r_k + \frac{1}{2}\left(\left(1-d\right)^2-\gamma'^2\right)\widetilde{ x_k^2}\right)
  -\left(\frac{\beta_{\textrm{r}}^2}{\gamma'}+\frac{\beta_{\textrm{r}}^2+2\beta_{\textrm{r}}}{\gamma'}\right)
  \left(\varphi_k-\left(1-d\right)\gamma'\widetilde{ x_k^2}\right)\\
  & = -2\left(\beta_{\textrm{r}}+1\right)r_k -\left(2+\beta_{\textrm{r}} + \frac{\beta_{\textrm{r}}^3+2\beta_{\textrm{r}}^2}{\gamma'^2}\right)
  \widetilde{r_k^2}\\
  & +\frac{1}{2}\left(\beta_{\textrm{r}}-\frac{\beta_{\textrm{r}}^3+2\beta_{\textrm{r}}^2}{\gamma'^2}\right)
  \left(\beta_{\textrm{r}}^2-\gamma'^2\right)\widetilde{ x_k^2}
  -2\left(\beta_{\textrm{r}}^2+\beta_{\textrm{r}}\right)\left(x_k+y_k\right)
  {-}2\left(\beta_{\textrm{r}}^2+\beta_{\textrm{r}}\right)\beta_{\textrm{r}}\widetilde{ x_k^2}\\
  & = -2\left(\beta_{\textrm{r}}+1\right)\left(-\beta_{\textrm{r}} x_k +\left(\beta_{\textrm{r}}+2\right)y_k\right)
  -\left(2+\beta_{\textrm{r}} + \frac{\beta_{\textrm{r}}^3+2\beta_{\textrm{r}}^2}{\gamma'^2}\right)
  \beta_{\textrm{r}}^2\widetilde{x_k^2}\\
  & +\frac{1}{2}\left(\beta_{\textrm{r}}-\frac{\beta_{\textrm{r}}^3+2\beta_{\textrm{r}}^2}{\gamma'^2}\right)
  \left(\beta_{\textrm{r}}^2-\gamma'^2\right)\widetilde{ x_k^2}
  -2\left(\beta_{\textrm{r}}^2+\beta_{\textrm{r}}\right)\left(x_k+y_k\right)
  {-}2\left(\beta_{\textrm{r}}^2+\beta_{\textrm{r}}\right)\beta_{\textrm{r}}\widetilde{ x_k^2}\\
  & = -4\left(\beta_{\textrm{r}}+1\right)^2y_k
  -\left(2+\beta_{\textrm{r}} + \frac{\beta_{\textrm{r}}^3+2\beta_{\textrm{r}}^2}{\gamma'^2}\right)
  \beta_{\textrm{r}}^2\widetilde{x_k^2}\\
  & +\frac{1}{2}\left(\beta_{\textrm{r}}-\frac{\beta_{\textrm{r}}^3+2\beta_{\textrm{r}}^2}{\gamma'^2}\right)
  \left(\beta_{\textrm{r}}^2-\gamma'^2\right)\widetilde{ x_k^2}
  {-}2\left(\beta_{\textrm{r}}^2+\beta_{\textrm{r}}\right)\beta_{\textrm{r}}\widetilde{ x_k^2}\\
  & = -4\left(\beta_{\textrm{r}}+1\right)^2y_k
  {-\left(4+3\beta_{\textrm{r}} + \frac{\beta_{\textrm{r}}^3+2\beta_{\textrm{r}}^2}{\gamma'^2}\right)}
  \beta_{\textrm{r}}^2\widetilde{x_k^2}\\
  & +\frac{1}{2}\left(\beta_{\textrm{r}}-\frac{\beta_{\textrm{r}}^3+2\beta_{\textrm{r}}^2}{\gamma'^2}\right)
  \left(\beta_{\textrm{r}}^2-\gamma'^2\right)\widetilde{ x_k^2}\\
  & = -4\left(\beta_{\textrm{r}}+1\right)^2y_k
       {-\left(4+\frac{5}{2}\beta_{\textrm{r}} + \frac{3\beta_{\textrm{r}}^3+6\beta_{\textrm{r}}^2}{2\gamma'^2}\right)
         \beta_{\textrm{r}}^2 \widetilde{x_k^2} - \frac{1}{2}\left(\beta_{\textrm{r}}\gamma'^2-\beta_{\textrm{r}}^3-2\beta_{\textrm{r}}^2\right)\widetilde{x_k^2}
       }\\
  & = -4\left(\beta_{\textrm{r}}+1\right)^2a\widetilde{x_k^2}
       {-\left(3+2\beta_{\textrm{r}} + \frac{3\beta_{\textrm{r}}^3+6\beta_{\textrm{r}}^2}{2\gamma'^2}\right)
         \beta_{\textrm{r}}^2 \widetilde{x_k^2} - \frac{1}{2}\beta_{\textrm{r}}\gamma'^2\widetilde{x_k^2}
       }\\
  & = -4\left(\beta_{\textrm{r}}+1\right)^2a\widetilde{x_k^2}
       {-\frac{\beta_{\textrm{r}}}{2\gamma'^2}\left(\gamma'^4+6\beta_{\textrm{r}}\gamma'^2+4\beta_{\textrm{r}}^2\gamma'^2 + 3\beta_{\textrm{r}}^4+6\beta_{\textrm{r}}^3\right) \widetilde{x_k^2}
       }\\
  & = -4\left(\beta_{\textrm{r}}+1\right)^2a\widetilde{x_k^2}
       {-\frac{\beta_{\textrm{r}}}{2\gamma'^2}\left(\gamma'^2+\beta_{\textrm{r}}^2\right)\left(3\beta_{\textrm{r}}\left(\beta_{\textrm{r}}+2\right)+\gamma'^2\right) \widetilde{x_k^2}
       }
\end{align*}
Similar to $R$, the $y_k$ are tangential to the center manifold. This translates into the fact that
\begin{equation*}
  \dot{y}_k = \left(\frac{\mathrm{d}}{\mathrm{d}x_k}y_k\right) \dot{x}_k
\end{equation*}
vanishes up to second order in $x_k$.
Therefore, comparing the terms in front of the $\widetilde{x_k^2}$ above yields 
\begin{align*}
  a &= {-\frac{\beta_{\textrm{r}}\left(\gamma'^2+\beta_{\textrm{r}}^2\right)\left(3\beta_{\textrm{r}}\left(\beta_{\textrm{r}}+2\right)+\gamma'^2\right)}{8\left(\beta_{\textrm{r}}+1\right)^2\gamma'^2}}\\
  & = {\frac{\left(1-d\right)\left(\gamma'^2+\left(1-d\right)^2\right)\left(3\left(d^2-1\right)
      +\gamma'^2\right)}{8d^2\gamma'^2}}.
\end{align*}

\subsection{$A$, $B$ and $C$}

Finally, the coefficients $A$, $B$ and $C$ for the dynamics in the center manifold,
cf. Eq.~\eqref{eq:cmf}, can be obtained by expanding the dynamics of $x_k$,
\begin{equation}
  2d \dot{x}_k = -\left(-1+\left(d+1\right)\frac{\gamma}{\gamma'}\right) e^{2R}\widetilde{e^{2r_k}}
  + \operatorname{Re} \left(\left(-1-i\frac{d+1}{\gamma'}\right)k\left(
  \langle e^{\mathbf{z}}\rangle \widetilde{e^{-z_k}}\right)\right)
  \tag{C.2}\label{eq:x}\,,
\end{equation}
in powers of $x_k$:
The terms in front of $\widetilde{x_k^2}$, $\widetilde{x_k^3}$ and $x_k \langle \mathbf{x}^2 \rangle$
correspond to the coefficients $A$, $B$ and $C$, respectively.
In order to do so, we approximate several terms as follows:
\begin{align*}
  \langle e^{\mathbf{z}}\rangle \widetilde{e^{-z_k}}
  & = -z_k + \frac{1}{2}\widetilde{z_k^2} - \frac{1}{6}\widetilde{z_k^3}
  -\frac{1}{2} z_k \langle \mathbf{z}^2\rangle  + \mathcal{O}\left(x_k^4\right)\\
  z_k & = \left(1-d+i\gamma'\right)x_k+a\left(1+d+i\gamma'\right)\widetilde{x_k^2}
  + \mathcal{O}\left(x_k^3\right)\\
  z_k^2 & = \left(1-d+i\gamma'\right)^2 x_k^2 +
  2a\left(1-d+i\gamma'\right)\left(1+d+i\gamma'\right)x_k\widetilde{x_k^2}+ \mathcal{O}\left(x_k^4\right)\\
  \widetilde{z_k^2} & = z_k^2-\langle \mathbf{z}^2 \rangle \\
  & = \left(1-d+i\gamma'\right)^2 \widetilde{x_k^2} +
  2a\left(1-d+i\gamma'\right)\left(1+d+i\gamma'\right)\left(x_k\widetilde{x_k^2}-\langle \mathbf{x}\,\widetilde{\mathbf{x}^2}\rangle\right) + \mathcal{O}\left(x_k^4\right)\\
  x_k\widetilde{x_k^2}-\langle \mathbf{x}\, \widetilde{\mathbf{x}^2}\rangle & = x_k^3 - x_k \langle \mathbf{x}^2 \rangle
  - \langle \mathbf{x}^3 \rangle + \langle \mathbf{x} \langle \mathbf{x}^2 \rangle\rangle\\
  & = \widetilde{x_k^3} - x_k \langle \mathbf{x}^2 \rangle\\
  \widetilde{z_k^2} & = \left(1-d+i\gamma'\right)^2 \widetilde{x_k^2} +
  2a\left(1-d+i\gamma'\right)\left(1+d+i\gamma'\right)\left(\widetilde{x_k^3} - x_k \langle \mathbf{x}^2 \rangle\right) + \mathcal{O}\left(x_k^4\right)\\
  z_k^3  & =\left(1-d+i\gamma'\right)^3x_k^3  + \mathcal{O}\left(x_k^4\right)\\
  \widetilde{z_k^3}  & =\left(1-d+i\gamma'\right)^3\widetilde{x_k^3}  + \mathcal{O}\left(x_k^4\right)\\
  z_k \langle \mathbf{z}^2\rangle & = \left(\left(1-d+i\gamma'\right)x_k+a\left(1+d+i\gamma'\right)\widetilde{x_k^2}\right) \\
  & \cdot
  \langle \left(1-d+i\gamma'\right)^2 \mathbf{x}^2 +
  2a\left(1-d+i\gamma'\right)\left(1+d+i\gamma'\right)\mathbf{x}\, \widetilde{\mathbf{x}^2}\rangle + \mathcal{O}\left(x_k^4\right)\\
  & = \left(\left(1-d+i\gamma'\right)x_k+a\left(1+d+i\gamma'\right)\widetilde{x_k^2}\right)
  \langle \left(1-d+i\gamma'\right)^2 \mathbf{x}^2 \rangle + \mathcal{O}\left(x_k^4\right)\\
  & = \left(1-d+i\gamma'\right)^3x_k\langle \mathbf{x}^2 \rangle + \mathcal{O}\left(x_k^4\right).
\end{align*}
Using these terms, we can write
\begin{align*}
  \langle e^{\mathbf{z}}\rangle \widetilde{e^{-z_k}}
  & = -z_k + \frac{1}{2}\widetilde{z_k^2} - \frac{1}{6}\widetilde{z_k^3}
  -\frac{1}{2} z_k \langle \mathbf{z}^2\rangle  + \mathcal{O}\left(x_k^4\right)\\
  & = -\left(1-d+i\gamma'\right)x_k - a\left(1+d+i\gamma'\right)\widetilde{x_k^2}\\
  & + \frac{1}{2}\left(1-d+i\gamma'\right)^2 \widetilde{x_k^2} +
  a\left(1-d+i\gamma'\right)\left(1+d+i\gamma'\right)\left(\widetilde{x_k^3} -
  x_k \langle \mathbf{x}^2 \rangle\right)\\
  &  - \frac{1}{6}\left(1-d+i\gamma'\right)^3\widetilde{x_k^3}\\
  &  - \frac{1}{2}\left(1-d+i\gamma'\right)^3x_k\langle \mathbf{x}^2 \rangle\\
  & = -\left(1-d+i\gamma'\right)x_k  \\
  & + \left(\frac{1}{2}\left(1-d+i\gamma'\right)^2-a\left(1+d+i\gamma'\right)\right)\widetilde{x_k^2}\\
  & + \left(a\left(1-d+i\gamma'\right)\left(1+d+i\gamma'\right)- \frac{1}{6}\left(1-d+i\gamma'\right)^3\right)
  \widetilde{x_k^3}\\
  & + \left(- \frac{1}{2}\left(1-d+i\gamma'\right)^3 - a\left(1-d+i\gamma'\right)\left(1+d+i\gamma'\right)\right)
  x_k \langle \mathbf{x}^2 \rangle
\end{align*}
\begin{align*}
  e^{2R}\widetilde{e^{2r_k}} & = \left(1+2b\langle \mathbf{x}^2\rangle\right)
  \left(2r_k + 2\widetilde{r_k^2}+\frac{4}{3}\widetilde{r_k^3}\right)+ \mathcal{O}\left(x_k^4\right)\\
  & = 2r_k + {4}br_k\langle \mathbf{x}^2\rangle + 2\widetilde{r_k^2}+\frac{4}{3}\widetilde{r_k^3}
  + \mathcal{O}\left(x_k^4\right)\\
  r_k & = \left(1-d\right) x_k + \left(1+d\right)y_k\\
  & = \left(1-d\right) x_k + \left(1+d\right)a \widetilde{x_k^2}\\
  r_k^2 & = \left(1-d\right)^2 x_k^2 + 2 a \left(1-d\right)\left(1+d\right)x_k\widetilde{x_k^2} +
  \mathcal{O}\left(x_k^4\right)\\
  r_k^3 & = \left(1-d\right)^3 x_k^3+\mathcal{O}\left(x_k^4\right)\\
  \widetilde{r_k^2} & = r_k^2 -\langle \mathbf{r}^2 \rangle\\
  & = \left(1-d\right)^2 \widetilde{x_k^2} + 2 a \left(1-d\right)\left(1+d\right)
  \left(\widetilde{x_k^3} - x_k \langle \mathbf{x}^2 \rangle\right)\\
  \widetilde{r_k^3} & = \left(1-d\right)^3 \widetilde{x_k^3}+\mathcal{O}\left(x_k^4\right)
\end{align*}
\begin{align*}
  e^{2R}\widetilde{e^{2r_k}}  & = 2r_k + {4}br_k\langle \mathbf{x}^2\rangle +
  2\widetilde{r_k^2}+\frac{4}{3}\widetilde{r_k^3}
  + \mathcal{O}\left(x_k^4\right)\\
  & = 2\left(1-d\right) x_k + 2a\left(1+d\right)\widetilde{x_k^2}\\
  & + {4}b\left(1-d\right) x_k\langle \mathbf{x}^2\rangle\\
  & + 2\left(1-d\right)^2 \widetilde{x_k^2} + 4 a \left(1-d\right)\left(1+d\right)
  \left(\widetilde{x_k^3} - x_k \langle \mathbf{x}^2 \rangle\right)\\
  & + \frac{4}{3}\left(1-d\right)^3 \widetilde{x_k^3}+\mathcal{O}\left(x_k^4\right)\\
  & = 2\left(1-d\right) x_k\\
  & + \left(2a\left(1+d\right)+2\left(1-d\right)^2\right)\widetilde{x_k^2}\\
  & + \left(4 a \left(1-d\right)\left(1+d\right) + \frac{4}{3}\left(1-d\right)^3\right)
  \widetilde{x_k^3}\\
  & + \left({4}b\left(1-d\right) - 4 a \left(1-d\right)\left(1+d\right)\right)x_k \langle \mathbf{x}^2 \rangle.
\end{align*}
We can now insert the different orders of $x_k$ from $e^{2R}\widetilde{e^{2r_k}}$ and
$\langle e^{\mathbf{z}}\rangle \widetilde{e^{-z_k}}$ in Eq.~\eqref{eq:x} (here, we use sympy~\cite{meurer17_sympy}
to solve for the coefficients), yielding
\begin{align*}
  2d\dot{x}_k & = \frac{\left(d-1\right)\left(\gamma'^2+\left(1+d\right)^2\right)\left(\gamma'^2-3\left(d-1\right)^2\right)}{2\gamma'^2}\widetilde{x_k^2}\\
  & -\frac{\left(d-1\right)^2\left(\gamma'^2+\left(d-1\right)^2\right)\left(\gamma'^2+\left(d+1\right)^2\right)\left(\gamma'^2-2\gamma'd+3\left(d^2-1\right)\right)\left(\gamma'^2+2\gamma'd+3\left(d^2-1\right)\right)}{8\gamma'^4d^2}\widetilde{x_k^3}\\
  & +\frac{\left(1-d\right)^2}{8d^2}\left( \vphantom{\int_1^2}\gamma'^4 - 4 \gamma'^2\left( 2d^3 - 7d^2 +1\right)\right.\\
  &- 2\left(8d^5 + d^4 - 56d^3 + 22 d^2 + 1\right)- \frac{4}{\gamma'^2}\left(2d^7 + 5d^6 - 4d^5 - 13d^4 + 2d^3 + 11d^2 - 3\right)\\
  &\left.+ \frac{9}{\gamma'^4}\left(d^2-1\right)^4 \vphantom{\int_1^2}\right)x_k\langle \mathbf{x}^2\rangle.
\end{align*}
Reading off the coefficients then gives the parameters
\begin{align*}
  A & = \frac{\left(d-1\right)\left(\gamma'^2+\left(1+d\right)^2\right)\left(\gamma'^2-3\left(d-1\right)^2\right)}{4\gamma'^2d}\\
  B & = -\frac{\left(d-1\right)^2\left(\gamma'^2+\left(d-1\right)^2\right)\left(\gamma'^2+\left(d+1\right)^2\right)\left(\gamma'^2-2\gamma'd+3\left(d^2-1\right)\right)\left(\gamma'^2+2\gamma'd+3\left(d^2-1\right)\right)}{16\gamma'^4d^3}\\
  C &= \frac{\left(d-1\right)^2}{16d^3\gamma'^4}\left(\vphantom{\int_1^2} 
  \gamma'^8 -4\gamma'^6\left(2d^3-7d^2+1\right)-2\gamma'^4\left(8d^5+d^4-56d^3+22d^2+1\right)\right. \nonumber\\
  & \left.-4\gamma'^2\left(2d^7+5d^6-4d^5-13d^4+2d^3+11d^2-3\right)+9\left(d^2-1\right)^4
  \vphantom{\int_1^2} \right).
\end{align*}

\section{2-Cluster states in the center manifold}
\label{sec:twocl_cmf}
For 2-cluster states, we can take $N=N_1+N_2$ and write
\begin{align*}
  \dot{x}_k &= \lambda_+ x_k + A\widetilde{x_k^2} + B\widetilde{x_k^3} + C\langle \mathbf{x}^2\rangle x_k
  + \mathcal{O}\left(x_k^4\right)\\
  & = \lambda_+ x_k + A\left(x_k^2-\frac{1}{N}\left(N_1x_1^2+N_2x_2^2\right)\right)+
  B\left(x_k^3-\frac{1}{N}\left(N_1x_1^3+N_2x_2^3\right)\right) + \frac{C}{N}\left(N_1x_1^2+N_2x_2^2\right)x_k
\end{align*}
with the constraint $k\in \left\{1, 2\right\}$ and $N_1x_1+N_2x_2=0$, that is,
$x_2=-(N_1/N_2)x_1$.
Note that $\dot{x}_k$ must vanish at the 2-cluster equilibria.
The 2-cluster therefore satisfies
\begin{align*}
  0 &= \lambda_+ x_1 + A\left(x_1^2-\frac{1}{N}\left(N_1x_1^2+\frac{N_1^2}{N_2}x_1^2\right)\right)+
  B\left(x_1^3-\frac{1}{N}\left(N_1x_1^3-\frac{N_1^3}{N_2^2}x_1^3\right)\right) + \frac{C}{N}\left(N_1x_1^2+\frac{N_1^2}{N_2}x_1^2\right)x_1\\
  & =\lambda_+ x_1 + A\left(x_1^2-\frac{N_1}{N_2}x_1^2\right)+
  B\left(x_1^3-\frac{N_1\left(N_2-N_1\right)}{N_2^2}x_1^3\right) + \frac{CN_1}{N_2}x_1^3\\
  & = \lambda_+ x_1 + A\frac{N_2-N_1}{N_2}x_1^2+
  B\frac{N_2^2-N_1\left(N_2-N_1\right)}{N_2^2}x_1^3+ \frac{CN_1}{N_2}x_1^3,
\end{align*}
and writing $\alpha=N_1/N_2$,
\begin{align*}
  0 & = \lambda_+ x_1 + A\left(1-\alpha\right)x_1^2+
  \left(B\left(1-\alpha+\alpha^2\right)+ C\alpha\right)x_1^3.
\end{align*}
This equation has the solutions $x_1=0, x_2=0$ and
\begin{align*}
  x_1^{\pm} & = \frac{1}{2\left(B\left(1-\alpha+\alpha^2\right)+ C\alpha\right)}
  \left(-A\left(1-\alpha\right)\pm\sqrt{A^2\left(1-\alpha\right)^2-4\lambda_+\left(B\left(1-\alpha+\alpha^2\right)+ C\alpha\right)}\right)\\
  x_2^{\pm}&=-(N_1/N_2)x_1^{\pm}.
\end{align*}
The saddle-node curves creating the 2-cluster solutions are thus parametrized by the vanishing
discriminant
\begin{align*}
  0 &= A^2\left(1-\alpha\right)^2-4\lambda_+\left(B\left(1-\alpha+\alpha^2\right)+ C\alpha\right)\\
  \Rightarrow \lambda_+ = \lambda_{\text{sn}} & = \frac{A^2\left(1-\alpha\right)^2}{4\left(B\left(1-\alpha+\alpha^2\right)+ C\alpha\right)}
\end{align*}
for unbalanced cluster solutions, that is, $\alpha\neq 1$ or $N_1\neq N_2$.
Thus, at the saddle-node bifurcation
\begin{equation*}
  x_1^{\pm} = x_1^{\ast} = -\frac{A\left(1-\alpha\right)}{2\left(B\left(1-\alpha+\alpha^2\right)+ C\alpha\right)}.
\end{equation*}
%% In order to show that $x_1^{\ast}$ is monotonic in $\alpha$, we calculate
%% \begin{align*}
%%   \frac{\partial}{\partial \alpha}x_1^{\ast} & =
%%   -\frac{-2A\left(B\left(1-\alpha+\alpha^2\right)+ C\alpha\right)
%%     -2A\left(1-\alpha\right)\left(B\left(2\alpha-1\right)+C\right)}
%%   {4\left(B\left(1-\alpha+\alpha^2\right)+ C\alpha\right)^2}\\
%% & =  A\frac{B\left(1-\alpha+\alpha^2\right)+ C\alpha
%%     +B\left(2\alpha-1\right)+C-\alpha\left(B\left(2\alpha-1\right)+C\right)}
%%   {2\left(B\left(1-\alpha+\alpha^2\right)+ C\alpha\right)^2}\\
%%   & = A\frac{B\left(2\alpha-\alpha^2\right)+C}
%%   {2\left(B\left(1-\alpha+\alpha^2\right)+ C\alpha\right)^2}.\\
%% \end{align*}
%% Note that the numerator is zero only if
%% \begin{equation*}
%% \alpha_0 = -\frac{1}{2B}\left(-2B\pm\sqrt{4B^2+4BC}\right) = 1\mp \frac{\sqrt{B^2+BC}}{B}
%% \end{equation*}
\bibliography{../../../ref/lit.bib}
\bibliographystyle{unsrt}
\end{document}